\documentclass[longauth]{aa} 

\usepackage{natbib,twoopt}
\usepackage[breaklinks=true,colorlinks=true,urlcolor=blue,linkcolor=blue,citecolor=blue,bookmarks=true,bookmarksopen=true]{hyperref}
\bibpunct{(}{)}{;}{a}{}{,} 					

\makeatletter
\newcommandtwoopt{\citeads}[3][][]{\href{http://adsabs.harvard.edu/abs/#3}%

{\def\hyper@linkstart##1##2{}%
\let\hyper@linkend\@empty\citealp[#1][#2]{#3}}}
\newcommandtwoopt{\citepads}[3][][]{\href{http://adsabs.harvard.edu/abs/#3}%
{\def\hyper@linkstart##1##2{}%
\let\hyper@linkend\@empty\citep[#1][#2]{#3}}}
\newcommandtwoopt{\citetads}[3][][]{\href{http://adsabs.harvard.edu/abs/#3}%
{\def\hyper@linkstart##1##2{}%
\let\hyper@linkend\@empty\citet[#1][#2]{#3}}}
\newcommandtwoopt{\citeyearads}[3][][]%
{\href{http://adsabs.harvard.edu/abs/#3}
{\def\hyper@linkstart##1##2{}%
\let\hyper@linkend\@empty\citeyear[#1][#2]{#3}}}
\makeatother
\usepackage{graphicx}
\usepackage[varg]{txfonts}
\usepackage{xcolor}

\newcommand{\derot}{$\theta_\mathrm{der}$}

\newcommand{\ff}{f\mbox{}f}

\newcommand{\ffi}{f\mbox{}f\mbox{i}}
\newcommand{\fix}{f\mbox{}i}
\newcommand{\fl}{f\mbox{}l}

\newcommand{\qp}{$Q_\phi$ }
\newcommand{\up}{$U_\phi$ }

\newcommand\T{\rule{0pt}{2.6ex}}
\newcommand\B{\rule[-1.2ex]{0pt}{0pt}}
\newcommand{\revia}{\rm}

\begin{document}

\title{The polarimetric imaging mode of VLT/SPHERE/IRDIS I:}
\subtitle{Description,  data reduction and observing strategy\thanks{
Based on observations made with ESO Telescopes at the La Silla Paranal Observatory under programme ID 095.C-0273(D)
}}
\titlerunning{The polarimetric imaging mode of VLT/SPHERE/IRDIS I}
\authorrunning{Jozua de Boer et al.}
\author{J.~de Boer \inst{1}
\and M.~Langlois \inst{2,3}
\and R.\,G.~van Holstein \inst{1,4}
\and J.\,H.~Girard \inst{5}
\and D.~Mouillet \inst{6}
\and A.~Vigan \inst{3}
\and K.~Dohlen \inst{3}
\and F.~Snik \inst{1}
\and C.\,U.~Keller \inst{1}
\and C.~Ginski \inst{1,7}
\and D.\,M.~Stam \inst{8}
\and J.~Milli \inst{4}
\and Z.~Wahhaj \inst{4}
\and M.~Kasper \inst{9}
\and H.\,M.~Schmid \inst{10}
\and P.~Rabou \inst{6}
\and L.~Gluck \inst{6}
\and E.~Hugot \inst{3}
\and D.~Perret \inst{11}
\and P.~Martinez \inst{12}
\and L.~Weber \inst{13}
\and J.~Pragt \inst{14}
\and J.-F. Sauvage \inst{15}
\and A.~Boccaletti \inst{11}
\and H.~Le Coroller \inst{3}
\and C.~Dominik \inst{7}
\and T.~Henning \inst{16}
\and E.~Lagadec \inst{12}
\and F.~Ménard \inst{6}
\and M.~Turatto \inst{17}
\and S.~Udry \inst{13}
\and G.~Chauvin \inst{6}
\and M.~Feldt \inst{16}
\and J.-L.~Beuzit \inst{3}
}

\institute{Leiden Observatory, Universiteit Leiden, 
P.O. Box 9513, 2300 RA Leiden, The Netherlands.\\
email: deboer@strw.leidenuniv.nl
\and CRAL, UMR 5574, CNRS, Université de Lyon, Ecole Normale Supérieure de Lyon, 46 Allée d’Italie, F-69364 Lyon Cedex 07, France
\and Aix Marseille Univ, CNRS, CNES, LAM, Marseille, France
\and European Southern Observatory (ESO), Alonso de Córdova 3107, Casilla 19001, Santiago, Chile.
\and Space Telescope Science Institute, 3700 San Martin Drive, Baltimore, MD 21218, USA
\and Univ. Grenoble Alpes, CNRS, IPAG, F-38000 Grenoble, France
\and Anton Pannekoek Instituut, Universiteit van Amsterdam, Science Park 904, 1098 XH Amsterdam, The Netherlands
\and Faculty of Aerospace Engineering, Delft University of Technology, Kluyverweg 1, 2629 HS Delft, The Netherlands
\and European Southern Observatory (ESO), Karl-Schwarzschild-Str. 2, 85748 Garching, Germany
\and ETH Zurich, Institute for Particle Physics and Astrophysics, Wolfgang-Pauli-Strasse 27,8093 Zurich, Switzerland
\and LESIA, CNRS, Observatoire de Paris, Université Paris Diderot, UPMC, 5 place J. Janssen, 92190 Meudon, France
\and Université C\^ote d’Azur, OCA, CNRS, Lagrange, France
\and Geneva Observatory, Univ. of Geneva, Chemin des Maillettes 51, 1290 Versoix, Switzerland
\and NOVA Optical Infrared Instrumentation Group, Oude Hoogeveensedijk 4, 7991 PD Dwingeloo, The Netherlands
\and ONERA – 29 avenue de la Division Leclerc, 92322 Chatillon Cedex, France
\and Max Planck Institute for Astronomy, K\"onigstuhl 17, 69117 Heidelberg, Germany
\and INAF–Osservatorio Astronomico di Padova, Vicolo dell’ Osservatorio 5, 35122, Padova, Italy
}
   \date{Received \today; Accepted TBD}

  \abstract
   {Polarimetric imaging is one of the most e\ff ective
techniques for high-contrast imaging and characterization 
   of protoplanetary disks, and {\revia has the potential to} become instrumental in the characterization of exoplanets.
   The Spectro-Polarimetric High-contrast Exoplanet REsearch (SPHERE) instrument installed on the Very Large Telescope contains the InfraRed Dual-band Imager and Spectrograph (IRDIS) with a dual-beam polarimetric imaging (DPI) mode,
which o\ff ers the capability to obtain linear polarization images at {\revia high contrast and resolution}.
   }
   {We aim to provide an overview of the polarimetric imaging mode of VLT/SPHERE/IRDIS 
	and study its optical design to improve observing strategies and data reduction.
   }
   {For $H$-band observations of TW\,Hydrae, we compare two data reduction methods {\revia that correct for instrumental polarization e\ff ects in di\ff erent ways}: 
a minimization of the 'noise' image ($U_\phi$), and a polarimetric-model-based correction method that we have developed, as presented in Paper\,II of this study.
   }
   {
	{\revia
	We use observations of TW\,Hydrae to illustrate the data reduction. 
	In the images of the protoplanetary disk around this star we detect variability in the polarized intensity and angle of linear polarization with pointing-dependent instrument con\fix guration.
}
   We explain these variations as instrumental polarization e\ff ects and correct for these e\ff ects using our {\revia model-based correction method}.
   }
   {The polarimetric {\revia imaging} mode of IRDIS {\revia has proven to be a very successful and productive} high-contrast polarimetric imaging system.
   However, the instrument performance is strongly dependent on the speci\fix c instrument con\fix guration.
   We suggest adjustments to future observing strategies to optimize polarimetric e\ffi ciency {\revia in \fix eld tracking mode} by avoiding unfavourable derotator angles. 
We recommend {\revia reducing} on-sky data with {\revia the pipeline called IRDAP that includes} the model-based correction method (described in Paper\,II)
to optimally account for the remaining telescope and instrumental polarization e\ff ects
   and to retrieve the true polarization state of the incident light.
   }


\maketitle

\section{Introduction}
\label{sec:intro}

\subsection{High-contrast and resolution imaging polarimetry}

	Imaging planets and protoplanetary disks in the visible and near-infrared (NIR) 
	requires the observer to 
	account for the large
	contrasts between bright stars and their faint surroundings.
	Polarimetry has proven to be a powerful tool for high-contrast imaging, 
	e.g. with HST/NICMOS \citep{Perrin:2009}, Subaru/HiCIAO \citep{Mayama:2012ApJ} and VLT/NACO \citep{Quanz:2011}.
	When starlight is scattered by circumstellar material it becomes polarized.
	Therefore it is possible to distinguish this scattered light from the predominantly unpolarized stellar speckle halo
	by computing the di\ff erence between two images recorded in two orthogonal polarization states. 
	This high-contrast imaging technique is known as Polarimetric Di\ff erential Imaging \citep[PDI; ][]{Kuhn:2001}.
	With the aid of  
	adaptive optics (AO), polarimetric imaging has been succesful in detecting faint circumstellar disks 
	down to very small separations \citep[${\sim}\,0.1''$; e.g.][]{2013ApJ...766L...2Q, garufi:2016}.
	Compared to alternative high-contrast imaging techniques such as Angular Di\ff erential Imaging \citep[ADI;][]{Marois:2006}, 
	PDI is especially well suited to image disks seen 
	close to face-on, 
	such as TW\,Hydrae 
	\citep[seen ${\sim}\,7^\circ$ from face-on orientation;][]{Rapson:2015,Boekel:2017}.
	While ADI suffers from self-subtraction of signal from a disk with a low inclination, 
	PDI remains sensitive to its scattered light.
	PDI will remove unpolarized stellar and disk signal alike, 
	{\revia which makes this technique only less suitable to detect}
	disks with very low degrees of polarization due to unfavorable scattering angles 
	(close to $0^\circ$ or $180^\circ$) or grain sizes much larger than the wavelength \citep{Hansen:1974}.
	Fortunately, the scattering surfaces of protoplanetary disks predominantly contain submicron sized grains (i.e. smaller than the typical wavelengths at which high-contrast imaging instruments operate),
	while the single-scattering angles in most regions of any circumstellar disk will not be close to $0^\circ$ or $180^\circ$.
	
	Apart from being an e\ff ective high-contrast imaging technique, 
	polarimetry o\ff ers the potential to characterize scattering particles in circumstellar disks and the atmospheres of exoplanets.
	Radiative-transfer modeling of disks is heavily plagued by degeneracies when the models are based on 
	the Spectral Energy Distribution (SED) alone \citep[e.g.][]{Andrews:2011, Dong:2012}. 
	\citet{Perrin:2015} and \citet{Ginski:2016} have used the resolved polarimetric surface brightness to determine the scattering phase 
	function for the debris disk around HR\,4796A {\revia \citep[see also][]{Milli:2015}} and HD\,97048, respectively.
	The scattering phase function will be instrumental in the unambiguous characterization of micron-sized dust particles.

Young self-luminous giant exoplanets or companion brown dwarfs can also be polarized at NIR wavelengths, as their thermal emission is scattered by cloud and haze particles in the companions'\,outer atmospheres or dust particles surrounding the companions~\citep{Sengupta:2010, Kok:2011, Marley:2011, Stolker:2017}. 
Substellar companions are observed as point sources, and only produce a signi\fix cant (integrated) polarization signal if the shapes of these companions projected on the image plane deviate from circular symmetry.
Measuring a polarization signal from a companion con\fix rms the presence of a scattering medium (e.g. clouds) and can trace the cloud morphology (e.g.~horizontal bands), rotational \fl attening, the projected spin-axis orientation and the shape and orientation of a disk around the companion. 

\subsection{The polarimetric mode of the VLT/SPHERE INfraRed Dual-band Imager and Spectrograph: IRDIS/DPI}
\label{sec:introirdis}

	In 2014, the Spectro Polarimetric High-contrast Exoplanet REsearch 
	\citep[SPHERE;][]{Beuzit:2019} instrument was commissioned at Unit Telescope 3 (UT3) of the Very Large Telescope (VLT).
	{\revia This instrument contains the extreme-AO system 
	SAXO \citep[SPHERE AO for eXoplanet Observation;][]{Fusco:2006,Fusco:2016}, that consists, among other components of a high-order deformable mirror (DM) with $41\,{\times}\,41$ actuators and a Shack-Hartmann wavefront sensor that can operate up to 1200\,Hz \citep{Fusco:2016}. The wavefront sensor records in the visible regime and performs best for stars of $R = 9 - 10$\,mag, where it typically yields a Strehl ratio of $\geq 90\%$. Still, up to the magnitude limit of $R = 14 - 15$\,mag the AO system improves the performance with a factor of ${\sim}\,5$ \citep{Beuzit:2019}.
	This extreme-AO system supports} three scienti\fix c subsystems: 
	the (visible-light) Zurich IMaging POLarimeter \citep[ZIMPOL;][]{Schmid:2018}, 
	the (NIR) Integral Field Spectrograph \citep[IFS;][]{Claudi:2008SPIE},
	and the (Near) InfraRed Dual-band Imager and Spectrograph \citep[IRDIS;][]{Dohlen:2008SPIE}.
	
	IRDIS is primarily designed to detect planets in di\ff erential imaging modes combined with pupil tracking, 
	where the telescope pupil remains \fix xed on the detector and the image rotates with the parallactic angle.
	This rotation of the image during observations
	allows the removal of the stellar speckle halo by performing ADI.
	A beam splitter ensures that the star is imaged twice on the detector.
	Wide-band, broad-band or narrow-band \fix lters can be inserted in a common \fix lter wheel upstream from the beamsplitter to allow what is called 'classical imaging'.
	Downstream from the beam splitter, another wheel is present
	with which we can introduce two di\ff erent \fix lters for the separate beams. 
	Observations in two di\ff erent color \fix lters allows the detection of planets (e.g. by observing 
	methane absorption in their atmosphere) with Dual-Band Imaging {\revia\citep[DBI; e.g., ][]{Rosenthal:1996, Racine:1999, Marois:2000, Vigan:2010}}.

	The inclusion of orthogonal {\revia linear} polarization \fix lters (polarizers) in this {\revia second} \fix lter wheel makes IRDIS a polarimeter. 
	In this Dual-beam Polarimetric Imaging \citep[hereafter DPI or IRDIS/DPI;][]{Langlois:2014} mode, a rotatable half-wave retarder is inserted in the common path of SPHERE to modulate between the linear polarization components.
	{\revia IRDIS/DPI is currently o\ff ered in \fix eld and pupil tracking. }
	The requirements for DBI contrast have provided high image quality from which also DPI bene\fix ts, in particular, 
	high image stability that is essential for coronagraphy \citep{Boccaletti2008SPIE, Carbillet:2011, Guerri:2011}, and most importantly 
	a very low di\ff erential wave-front error between the two beams \citep{Dohlen:2016}.
	{\revia Since IRDIS/DPI was \fix rst o\ff ered to the community during the science veri\fix cation of SPHERE in december 2014, it has proven to be a very succesful and productive mode for} 
high-contrast imaging of circumstellar disks \citep[e.g.,][]{Benisty:2015, Stolker:2016, Garufi:2017,Avenhaus:2018, Pinilla:2018}, 
	but {\revia it} also shows great promise for the characterization of polarized substellar companions. 			
	{\revia \citet{Holstein:2017} have used IRDIS/DPI to search for a polarization signal in the companions around HR\,8799 and PZ\,Tel, similar to the attempts made with GPI for HD\,19467\,B by \citet{Jensen:2016} and $\beta$\,Pic\,b by \citet{Millar:2015}.
	Although \citeauthor{Holstein:2017} do not detect a polarization signal for the companions of either star, they do present stringent upper limits on the polarization of ${\sim} 0.1\%$ for PZ\,Tel\,B, and ${\sim} 1\%$ for the much fainter planets around HR\,8799. Furthermore they present a polarized contrast of ${\sim}\,10^{-7}$ at 0.5" separation from the primary star HR\,8799.}
	
	Due to the complexity of the SPHERE instrument and its many re\fl ecting surfaces, the polarimetric performance 
	is strongly dependent on the speci\fix c instrumental setup. 
	Each optical component in the telescope and instrument can cause \textit{instrumental polarization e\ff ects}, 
	which we group in two categories:
 	1) the {introduction} of polarization, and 2) the {mixing} of polarization states in the light beam, 
which we call \textit{Instrumental Polarization} ({\revia hereafter \textit{IP}, to avoid confusion with the overarching term "instrumental polarization effects"}) and polarimetric \textit{crosstalk}, respectively.
	\textit{IP} can give the false impression of a detection of polarization where there is in fact no true polarization signal incident on the telescope. 
	Crosstalk can change incident polarization into a state that is not being measured by the instrument (e.g., linear to circular polarimetry, see Sect.\,\ref{sec:pdi}). 
	Therefore, crosstalk can decrease the \textit{polarimetric e\ffi ciency}: the fraction of measured polarization over the incident polarization.
	Unlike the polarimetric mode of ZIMPOL, no hard requirements where de\fix ned for the polarimetric performance of IRDIS, because the initial science priority of this mode was determined to be low.
 	Lower instrumental polarization e\ff ects were expected in the NIR than in the visible. 
	Furthermore, because of the di\ffi culty of the system analysis required to predict and correct these e\ff ects beforehand, 
	the choice was made to rely on a-posteriori characterisation of the DPI mode, whenever possible.

	This work forms Paper\,I of a larger study describing the polarimetric imaging mode of SPHERE/IRDIS.
	In Paper\,I we will present an overview: we will focus on the description of the instrument, data reduction, and we will make sugestions for observing strategies to maximize the polarimetric 
	performance of the instrument in \fix eld-tracking mode.
	Field tracking, where the image of the star remains \fix xed on the detector, is the default mode for DPI and therefore the tracking mode we will use for this paper. 
	In Paper\,II (van Holstein et al. in prep.) we will describe the polarimetric instrument model that we have developed, based on 
	calibration measurements using unpolarized stars and SPHERE's internal light-source.
	Furthermore, in Paper\,II we will describe a correction method based on this model to account for the instrumental polarization e\ff ects and compute the true polarization signal incident on the telescope. {\revia This correction method is included in a new data-reduction pipeline called IRDIS Data reduction for Accurate Polarimetry (IRDAP), which we will make public (see Paper\,II}).

	Paper\,I begins with a general description of polarization and {\revia dual-beam} polarimetric imaging in Sect.\,\ref{sec:pdi}.
	We will describe the optical components encountered by the light beam 
	in Sect.\,\ref{sec:design}.
	{\revia In Sect.\,\ref{sec:postproc}, we will explain the basic principles behind the data reduction, which we will apply in Sect.\,\ref{sec:twh} on} the TW\,Hydrae observations of \citet{Boekel:2017}. 
	In the reduced data of TW\,Hydrae we detect an instrument-con\fix guration-dependent variation in the polarization signal, which we will use to illustrate the {\revia polarimetric} performance of {\revia IRDIS/DPI. 
	In the remainder of Sect.\,\ref{sec:tempvar} we will describe} the instrumental polarization e\ff ects {\revia of SPHERE/IRDIS with the use of} the polarimetric instrument model of Paper\,II. 
{\revia These instrumental polarization e\ff ects will enable us to explain the instrument-con\fix guration-dependent variations in TW\,Hydrae}.
	In Sect.\,\ref{sec:comparemeth} we will apply the correction method described in Paper\,II to account for the instrumental polarization e\ff ects and obtain the true polarization state for TW\,Hydrae. 
	We will compare the results of the {\revia IRDAP reduction, (with the model-based} correction method) with the results of the best {\revia 'conventional'} data reduction, 
{\revia where we apply an empirical} correction method. 
Based on our analysis, we will propose recommendations for future observations and possible SPHERE upgrades to enhance the polarimetric performance in Sect.\,\ref{sec:recommend}. 
{\revia In Sect.\,\ref{sec:competition} we will make a comparison between SPHERE/IRDIS/DPI and major contemporary AO-assisted polarimetric imagers in the NIR}.
We will end Paper\,I with our conclusions and recommendations in Sect.\,\ref{sec:conclude}.

\section{Dual-beam Polarimetric Imaging}
\label{sec:pdi}

\subsection{Polarization conventions and de\fix nitions}
\label{sec:pol101}

\begin{figure}[!h]
   \centering
   \includegraphics[width=0.5\textwidth, trim = 0 0 0 0]{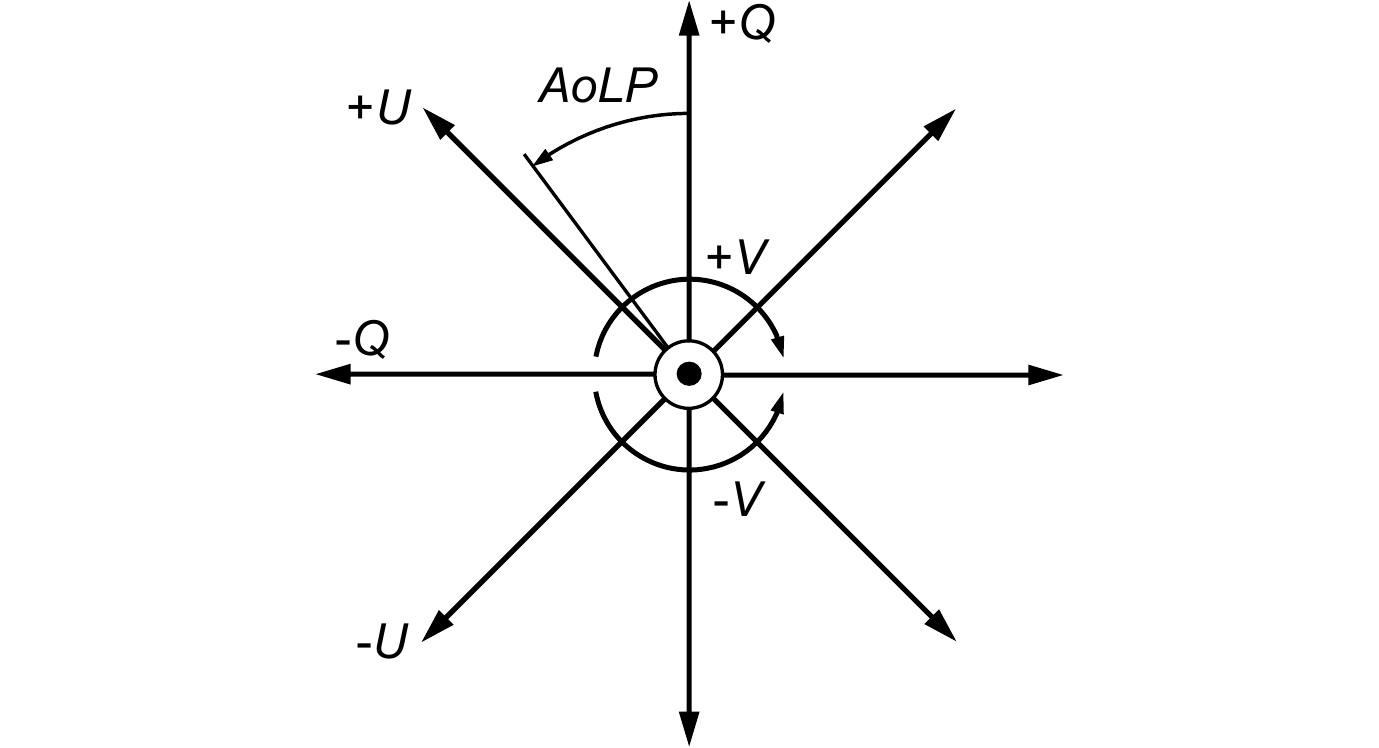}
   \caption{Orientation of the Stokes vector components of Eq.\,\ref{eq:stokes}, where the beam is propagating out of the paper towards the reader. After the vertical axis ($+Q$) is aligned with a prefered orientation (usually the meridian on-sky), the remaining Stokes vector components are oriented accordingly. The Angle of Linear Polarization ($AoLP$, Eq.\,\ref{eq:aolp}) is also measured with respect to the $+Q$ direction in counter-clockwise direction.
      \label{fig:poldef}}
    \end{figure}

Elliptical polarization (partial and full) is conveniently described by \citet{Stokes:1851} with what is known as the Stokes vector:
\begin{equation}
	\boldsymbol{S} = \left[ 
		\begin{array}{c}
			I \\
			Q \\
			U \\
			V
		\end{array}
		\right],
		\label{eq:stokes}
\end{equation}
where $I$ is the total intensity of the beam of light; 
$Q$ and $U$ describe the two linear polarization contributions; and $V$ describes circular polarization.
In the literature, the $+Q$ direction is often aligned with the local meridian \citep[e.g.,][]{Witzel11}.
In Sect.\,\ref{sec:postproc} we have used this convention for $+Q$. Although this choice of reference frame is arbitrary, {\revia it is the convention adopted by the International Astronomical Union}.
As illustrated in Fig.\,\ref{fig:poldef}, we have used the following conventions for the remaining orientations: 
For a beam propagating along the $z$-axis (in the direction of increasing $z$) in a right-handed $x,y,z$ coordinate system, let $+Q$ describe linear polarization with a prefered oscillation in the $ \pm x$ direction
(which we align with our frame of reference, e.g., the meridian);
$-Q$ then oscillates in the $\pm y$ direction; $+U$ describes linear polarization oscillating at an angle of $+45^\circ$ from the $x$-axis (rotated in counter-clockwise direction when looking at the source, and $-45^\circ$ from the $y$-axis); while $-U$ polarized light oscillates at an angle of $-45^\circ$  (or $+135^\circ$) with respect to the $x$-axis.
When the observer is facing towards the $-z$ direction,
$+V$ describes circular polarization where the peak of the electric \fix eld rotates clockwise  (i.e. moving from $+x$ to $-y$)
and $-V$ describes counter-clockwise rotation.

IRDIS/DPI is designed to measure linear polarization only, which is expected to be the dominant polarization component caused by scattering at the surface layers of protoplanetary disks and substellar companions.
From the Stokes vector components, we can determine the linearly Polarized Intensity ($PI_\mathrm{L}$); Degree and Angle of Linear Polarization ($DoLP$ or $P_\mathrm{L}$ \& $AoLP$, respectively) according to:
\begin{eqnarray}
	PI_\mathrm{L}    &=& \sqrt{Q^2 + U^2}, \label{eq:polint}\\
	 P_\mathrm{L}    &=& \frac{PI_\mathrm{L}}{I}  = \frac{\sqrt{Q^2 + U^2}}{I}, \label{eq:dolp}  \\
	{AoLP} &=& \frac{1}{2}\arctan{\left(\frac{U}{Q}\right)}. \label{eq:aolp}
 \end{eqnarray}

\subsection{The polarimetric imager}
\label{sec:pdi2}
Although ideal polarimeters do not exist,
such a hypothetical instrument is helpful when we describe the general principles of PDI.
Apart from mirrors and lenses the main components of this ideal polarimeter are two (in case of dual-beam) analyzers and detectors (or detector halves).
The analyzers can either be two separate polarizers (which require an aditional, preferably non-polarizing beamsplitter upstream) with orthogonal polarization (also called transmission) axes or a polarizing beamsplitter.  
Let us choose the 
{\revia polarization axis of one analyzer (A1) to be aligned with the $+Q$ direction, and the other analyzer (A2) to be aligned with $-Q$}.
We can retrieve (or `indirectly measure') the \fix rst two components of Eq.~\ref{eq:stokes} by adding and subtracting the measured intensity of both beams ($I_\mathrm{A1/A2}$) of light, respectively:
\begin{eqnarray}
	I_\mathrm{}   &=& I_\mathrm{A1} + I_\mathrm{A2}, 
	\label{eq:imeas}\\
	Q_\mathrm{} &=&  I_\mathrm{A1} - I_\mathrm{A2}. 
	\label{eq:qmeas}
\end{eqnarray}
We can rephrase Eqs.~\ref{eq:imeas} and \ref{eq:qmeas} to describe the transmission of the analyzers: 
\begin{eqnarray}
	I_\mathrm{A1} &=& \dfrac{1}{2}(I_\mathrm{} + Q_\mathrm{}), \label{eq:det1}\\
	I_\mathrm{A2} &=& \dfrac{1}{2}(I_\mathrm{} - Q_\mathrm{}),  \label{eq:det2}	
\end{eqnarray}
where for an ideal polarimeter, $I$ and $Q$ will be equal to
their counterparts incident on the telescope ($I_\mathrm{in}$ and $Q_\mathrm{in}$, respectively).

\begin{figure*}[!h]
   \centering
   \includegraphics[width=\textwidth, trim = 0 0 0 0]{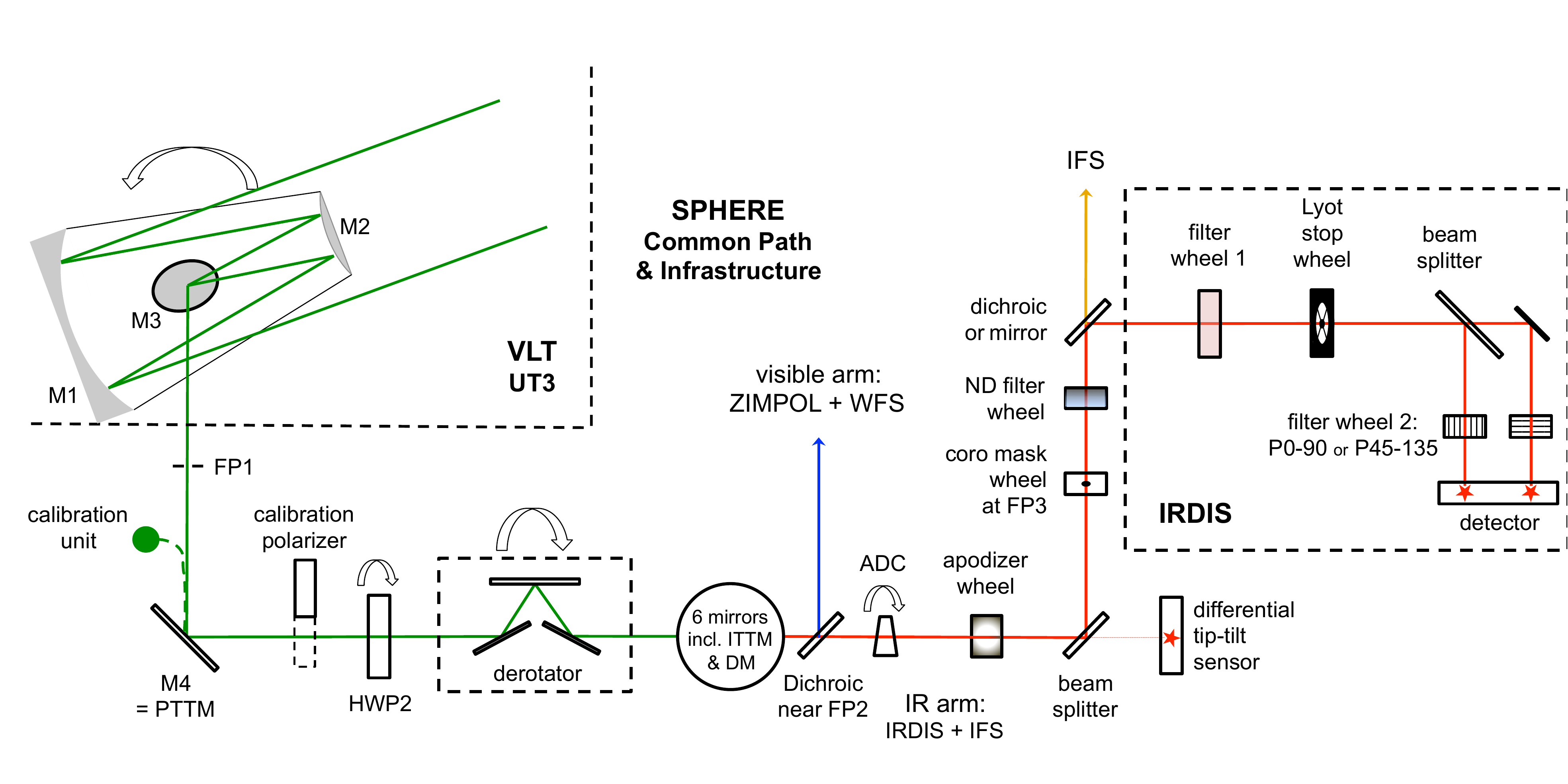}
   \caption{
	Schematic overview of the telescope and SPHERE/IRDIS, showing the optical components that are relevant for the polarimetric imaging mode.
	The curved arrows indicate components that rotate during an observation block. 
	Re\fl ections at angles of incidence $\geq 45^\circ$ in the instrument are represented
	with similarly large incidence re\fl ections in the figure.
	The green beam shows the starlight before color \fix lters are applied, blue represents visible light,
	red and orange represent NIR light (with the orange beam towards the IFS showing the shorter wavelengths).
      \label{fig:vltsphere}}
    \end{figure*}
	
	To retrieve $U$, we will either need to rotate the analyzers by $45^\circ$ or introduce an optical 
	component that can rotate the polarization direction with the same angle. 
	A half-wave ($\lambda/2$) plate (HWP) retards light that is polarized in the direction orthogonal to its 
	fast axis with $\lambda/2$ compared to light that is polarized in alignment with its fast axis.
	Therefore, a HWP upstream from the beam splitter can be used to rotate the measured polarization angle by $\Delta {AoLP}$ by placing the fast axis of 
	the HWP at an angle of $\Delta {AoLP}/2$ with respect to the polarization axes of the 
	analyzers \citep{Appenzeller:1967}.
	It is possible to retrieve $U$ by placing the HWP at an angle $\theta_\mathrm{HWP} = 22.5^\circ$ with 
	respect to the polarization axis of A1,
	which changes Eqs.~\ref{eq:det1} and \ref{eq:det2} into:
	$I_\mathrm{A1/A2} = (I \pm U)/2$, and Eq.~\ref{eq:qmeas} will yield $U$, {\revia instead of $Q$}. 
	We now see why the Stokes vector notation is convenient: 
	its components are easilly retrieved from the observables of an ideal polarimeter, 
	which `measures' a Stokes vector unaltered by the telescope and instrument 
	(i.e. $\boldsymbol{S} = \boldsymbol{S}_\mathrm{in}$, where $ \boldsymbol{S}_\mathrm{in}$ is the incident Stokes vector).
	
	Real polarimeters are never ideal: instrumental polarization e\ff ects depend on the speci\fix c instrument 
	con\fix guration used during observations.
	For complex instruments, the major instrumental polarization e\ff ect is typically the introduction of {\revia \textit{IP}}
	caused by the large number of re\fl ections in the telescope and instrument.
	We can correct for any {\revia \textit{IP}} created downstream of the HWP 
	by recording $Q$ for two HWP angles: $\theta_\mathrm{HWP} = 0^\circ\,\&\,45^\circ$
	\citep[see e.g.][]{1996aspo.book.....T, Witzel11, 2011A&A...531A.102C}.
 	The second $\theta_\mathrm{HWP}$ changes the signs 
	of the beam's original $Q$ component but leaves the \textit{IP} created downstream from the HWP unaltered.
	Therefore, for non-ideal polarimeters we change the notation {\revia of} the 
	\textit{single-di\ff erence} computations described by Eq.~\ref{eq:qmeas} to measure 
	$Q^+ = Q + IP$ for $\theta_\mathrm{HWP} = 0^\circ$, and $Q^- = -Q + IP$ for $\theta_\mathrm{HWP} = 45^\circ$.
	Similarly, we rename the \textit{single-sum} total intensities determined with Eq.\,\ref{eq:imeas} for $\theta_\mathrm{HWP} = 0^\circ$ \& $45^\circ$ as $I_{Q^+}$ \& $I_{Q^-}$, respectively.
	We then apply the \textit{double-di\ff erence} method
	to obtain the linear Stokes parameters corrected for {\revia \textit{IP}} created downstream of the HWP
and the corresponding total-intensity images with the \textit{double sum}:
	\begin{eqnarray}
		Q_\mathrm{} &=& \dfrac{1}{2}\left(Q_\mathrm{}^+ - Q_\mathrm{}^-\right) \label{eq:qddif}, \\
		I_Q &=& \dfrac{1}{2}\left(I_{Q^+} + I_{Q^-}\right) \label{eq:iqddif}, \\
		U_\mathrm{} &=& \dfrac{1}{2}\left(U_\mathrm{}^+ - U_\mathrm{}^-\right) \label{eq:uddif}, \\
		I_U &=& \dfrac{1}{2}\left(I_{U^+} + I_{U^-}\right) \label{eq:iuddif}, 
	\end{eqnarray}
	where $U^+= U + IP$ and $I_{U^+}$ are measured with $\theta_\mathrm{HWP} = 22.5^\circ$, 
	while $\theta_\mathrm{HWP} = 67.5^\circ$ yields $U^-= -U + IP$ and $I_{U^-}$.

The double di\ff erence does not remove \textit{IP} caused by the telescope and instrument mirrors upstream from the HWP, nor does it remove the most important crosstalk contributions. 
Correcting for these instrumental polarization e\ff ects requires that we determine the polarimetric response function for the
{\revia polarimetric imager, as we do in Sect.\,\ref{sec:intromodel} for} 
the polarimetric mode of VLT/SPHERE/IRDIS.

\section{Design of the polarimetric mode IRDIS/DPI}
\label{sec:design}

{\revia In this Section} we describe the optical components of {\revia VLT/UT3}, SPHERE's Common Path and Infrastructure (CPI) and IRDIS that are most important because they either create
	instrumental polarization e\ff ects, 
	they are useful for calibrations or they can be changed to modify the observational sequence or strategy.
	These optical components are illustrated in the schematic overview of the telescope and instrument in Fig.\,\ref{fig:vltsphere}.
	{\revia Especially re\fl ections at high angles of incidence are highlighted because}	
	larger angles
	are more prone to introduce instrumental polarization e\ff ects. 

\subsection{Telescope and SPHERE common path \& infrastructure}
\label{sec:UTCPI}

	SPHERE is installed on the Nasmyth platform of {\revia the alt-azimuth} {\revia Unit} Telescope 3. 
	After the axi-symmetric (and therefore non-polarizing) re\fl ections of the primary and secondary mirrors 
	(M1 and M2), the third mirror (M3) of UT3 is used to direct the light towards the Nasmyth focus.
	M3 introduces the \fix rst re\fl ection that breaks axi-symmetry {\revia with} a $45^\circ$\,angle of incidence.

	Shortly after the beam enters SPHERE we reach Focal Plane 1 (FP1), where a calibration light source
	can be inserted. 
	The \fix rst re\fl ection in the light path within SPHERE is the
	pupil tip-tilt mirror (PTTM or M4, with a $45^\circ$\,incidence angle), which is the only mirror in SPHERE {\revia that is} coated with 
	aluminum.\footnote{ 
	This coating gives M4 similar re\fl ective properties as M3 of UT3, 
	which is most useful for ZIMPOL. 
	SPHERE contains a visible light HWP (HWP1) between M3 and M4 that keeps the angles of the polarization induced by M3 crossed with that induced by M4, 
	e\ff ectively canceling both their contributions.
	Unfortunately, for the NIR, there is no HWP1 installed in SPHERE.
	This was chosen such because the instrumental polarization e\ff ects of these mirrors are smaller in the NIR and the 
	requirements for IRDIS' polarimetric performance less stringent than for ZIMPOL.
	} 
	The remaining mirrors of SPHERE are all coated with protected silver for its higher re\fl ectivity.
	A calibration polarizer with a \fix xed polarization angle can be inserted in the light path, 
	just before the beam encounters HWP2, the only HWP available for IRDIS/DPI.
	{\revia In \fix eld-tracking mode,} HWP2 can be rotated for two reasons. 
	The \fix rst reason is to switch between four angles 
	(HWP switch angles $\theta_\mathrm{HWP}^s = 0^\circ, 45^\circ, 22.5^\circ$ and $67.5^\circ$, where the superscript 's' is used to distinguish between switch angles and the true angle of HWP2) to measure 
	$Q^\pm$ and $U^\pm$ with IRDIS. 
	The second reason is to account for \fix eld rotation in order to keep the source polarization angle \fix xed relative to the analyzer for a given HWP2 switch angle in a polarimetric cycle.
	The next optical component downstream is the image derotator, composed of three mirrors {\revia that together form a `K'-shape (therefore also called} K-mirror, with the subsequent angles of incidence of $55^\circ$, $20^\circ$ and $55^\circ$). 
	The derotator rotates around the optical axis to stabilize either the \fix eld or the pupil on the detector.
	{\revia In Appendix\,\ref{sec:tracklaw}, we have included a detailed description of the tracking laws for HWP2 and the derotator in both \fix eld-stabilized and pupil-stabilized mode.}

	Multiple re\fl ective surfaces with small angles of incidence follow in the AO common path, 
	including the image tip-tilt mirror (ITTM), the $41\,{\times}\,41$ actuator {\revia high-order} DM and 
	three toric mirrors \citep{Hugot2012A&A}.				
	A dichroic beam splitter separates the light into a visible and a NIR arm just after Focal Plane 2 (FP2).
	Note that an additional focal plane exists between FP1 and FP2. 
	However, we adopt the nomenclature commonly used in existing literature such as the SPHERE manual.
	The visible light is re\fl ected by the dichroic beam splitter and sent to SAXO's 
	wavefront sensor and when required also to ZIMPOL 
	(currently not o\ff ered simultaneously with IRDIS or IFS).
	The NIR beam is transmitted by the dichroic beam splitter, and is then corrected for atmospheric dispersion, 
	determined by the airmass during the observations.
	Due to the low angles of incidence (${\leq}\,2.17^\circ$) on the prisms of the Atmospheric Dispersion Corrector (ADC), it is assumed not to cause signi\fix cant instrumental polarization e\ff ects. 
	The validity of this assumption is left for future investigation. 
	The beam then goes through the apodizer wheel (which allows apodization of the pupil in combination 
	with Lyot coronagraphs) and is sent to a beamsplitter, which transmits
	$2\%$ of the $H$ band to a di\ff erential tip-tilt sensor (DTTS) and re\fl ects the remaining light at $45^\circ$\,angle of incidence. 
	The (main) re\fl ected beam then encounters the wheel containing NIR coronagraph (focal) masks 
	\citep{Boccaletti2008SPIE, Martinez:2009}
	 in FP3, and the Neutral Density (ND) \fix lter wheel before reaching the \fix nal 
	 $45^\circ$\,angle re\fl ection that directs the beam towards IRDIS. 
	For this re\fl ection a dichroic beam splitter is selected when we use IRDIS in concert with IFS. 
	For modes that only use IRDIS, which is currently the case for polarimetry, a mirror is selected instead.

 \subsection{SPHERE/IRDIS}
 \label{sec:desirdis}
 	
 	IRDIS is described in detail by \citet{Dohlen:2008SPIE}. 
 	In this subsection and in Fig.\,\ref{fig:vltsphere}, we summarize the optical components for a better understanding of the polarimetric performance of the system and for reference later in this paper. 
	The optical components of IRDIS are located within a cryostat cooled to 100\,K to reduce thermal background emission.
 	The \fix rst optical component inside IRDIS is a common \fix lter wheel (FW1). 
The \fix lters of FW1 are the only color \fix lters that we can insert for the polarimetic mode, since FW2 contains the analyzer/polarizer pair.
 	{\revia Besides} narrow-band and spectroscopy \fix lters, FW1 contains four broad-band \fix lters which are o\ff ered for DPI (see Table~\ref{tab:filters}).
 
\begin{table}[!h]
\centering
\caption{
	Central wavelength ($\lambda_c$), bandwidth ($\Delta \lambda$) {\revia and pixel scale} for 
	SPHERE/IRDIS broad-band \fix lters available in \fix lter wheel 1. {\revia The wavelength and bandwidth are} described on the
	\href{https://www.eso.org/sci/facilities/paranal/instruments/sphere/inst/filters.html}{ESO website}, the pixel scales come from \citet{maire:2016a, maire:2018}. 
	The broad-band \fix lters are listed here as BB\_$X$, similar to ESO Observing Blocks (OBs), 
	but they are frequently listed as B\_$X$ (on the ESO website and in the headers of FITS files under keyword 
	INS1.FILT.NAME),  and as $X$ band in the main text.
{\revia $^{a)}$The pixel scale for BB\_$Y$ has not been calibrated. Therefore, the value for the $Y2$ \fix lter from \citet{maire:2016a} has been adopted.}
} 
\centering 
\begin{tabular}{l c c c} 
\hline\hline
\T\B Filter & $\lambda_c$ (nm) & $\Delta \lambda$ (nm) & {\revia Pixel scale (mas/pix)} \\ 
\hline
\T BB\_$Y$ & 1043 & 140 & $12.283^a\,\pm\,0.009$\\
BB\_$J$ & 1245 & 240  & $12.263\,\pm\,0.009$ \\
BB\_$H$ & 1625 & 290 & $12.251\,\pm\,0.009$  \\
\B BB\_${K_s}$ & 2182 & 300  & $12.265\,\pm\,0.009$ \\
\hline 
\end{tabular}  
\label{tab:filters}
\end{table}

 	Next, the beam encounters a Lyot stop wheel that also includes a mask for the pupil obscuration by M2 and its support structure (the "spider").
	Directly downstream from the Lyot stop wheel, the beam is split by the non-polarizing beam splitter plate (NBS).
 	The beam transmitted by the NBS is re\fl ected by an extra mirror in the direction parallel with the beam re\fl ected by the NBS (and therefore with the same angle of incidence as the re\fl ected beam: $45^\circ$).
 	The beams are \fix nally re\fl ected by two identical spherical camera mirrors, which focus the beams on the detector (not shown in Fig.\,\ref{fig:vltsphere}).
	
	The second \fix lter wheel (FW2) hosting the wire-grid polarizer pairs (P0-90 and P45-135) is located between the camera mirrors and the detector.	    
Finally, the beam reaches the Hawaii-2RG detector, which is mounted on a dither stage and has $2048\,{\times}\,2048$ pixels with 18 $\muup$m pitch. 
Each of the two orthogonally polarized beams is focused on a separate quadrant ($1024\,{\times}\,1024$ pixels) of the detector, which {\revia results in a \fix eld of view (FOV) of ${\sim}\,11''\,{\times}\,12.5''$, and the \fix lter-dependent pixel scales listed in table\,\ref{tab:filters} \citep{maire:2016a, maire:2018}.}


\subsection{Wire-grid polarizer pairs and beam splitter}
\label{sec:analyzers}
The P0-90 analyzer set in FW2 \fix lters the light with polarization angles perpendicular to and aligned with the plane of the Nasmyth platform: the plane in which all re\fl ections downstream from the derotator occur.
The P45-135 set polarizes at angles of $45^\circ$ and $135^\circ$ with respect to this plane.
Measurements recorded with P45-135 are highly sensitive to crosstalk introduced by all re\fl ections in this plane.
Therefore, we limit the study in Papers\,I\,\&\,II to the use of the P0-90 analyzer pair, while using HWP2 to switch between $Q^\pm$ and $U^\pm$ measurements, which is the default setup for DPI. 

	The non-polarizing beam splitter is not perfectly non-polarizing, which is corrected for when we use the double di\ff erence.
Therefore we can use the \fix rst-order approximation that it does not introduce new polarization to the beam.
Astrophysical objects in the \fix eld of high-contrast imaging typically have a very low degree of polarization when integrating the total beam, since this beam is dominated by the central predominantly unpolarized star. 
The polarizers will only transmit one polarization state each, which means that both beams will loose ${\sim}\,50\%$ of {\revia their} photons.


%

%
%
%

\section{Polarimetric data reduction}
\label{sec:postproc}

{\revia In this Section, we describe the basic steps of the polarimetric data reduction for the IRDIS/DPI mode. 
In Sect.\,\ref{sec:tempvar} we will apply this to the data of TW\,Hydrae as published by \citet{Boekel:2017}, 
and use the results to analyse the polarimetric performance of the system. 
Because these observations were recorded before we had performed polarimetric calibrations, 
we encountered unexpected instrumental polarization e\ff ects that depend on the speci\fix c instrument con\fix guration, and vary during the observing sequence. 
We will describe and explain these effects in detail based on the polarimetric instrument model of Paper\,II. 
In Sect.\,\ref{sec:comparemeth} we compare the reduction of this data after a data-driven correction for instrumental polarization e\ff ects with a data reduction after a correction based on the polarimetric instrument model}.
	
	{\revia To promote general understanding of the underlying principles of the polarimetric data reduction and because our analysis of the data in the subsequent sections} required non-standard tests, 
	we did not use the o\ffi cial 
	Data Reduction and Handling \citep[DRH,][]{Pavlov:2008} pipeline but our own custom data reduction routines described below.
	However, the pre-processing (background subtraction, \fl at \fix elding and centering) is very similar to what is done by DRH and therefore described in Appendix~\ref{sec:reduce}.

\begin{figure*}[!h]
   \centering
   \includegraphics[width=0.8\textwidth, trim = 0 0 0 10]{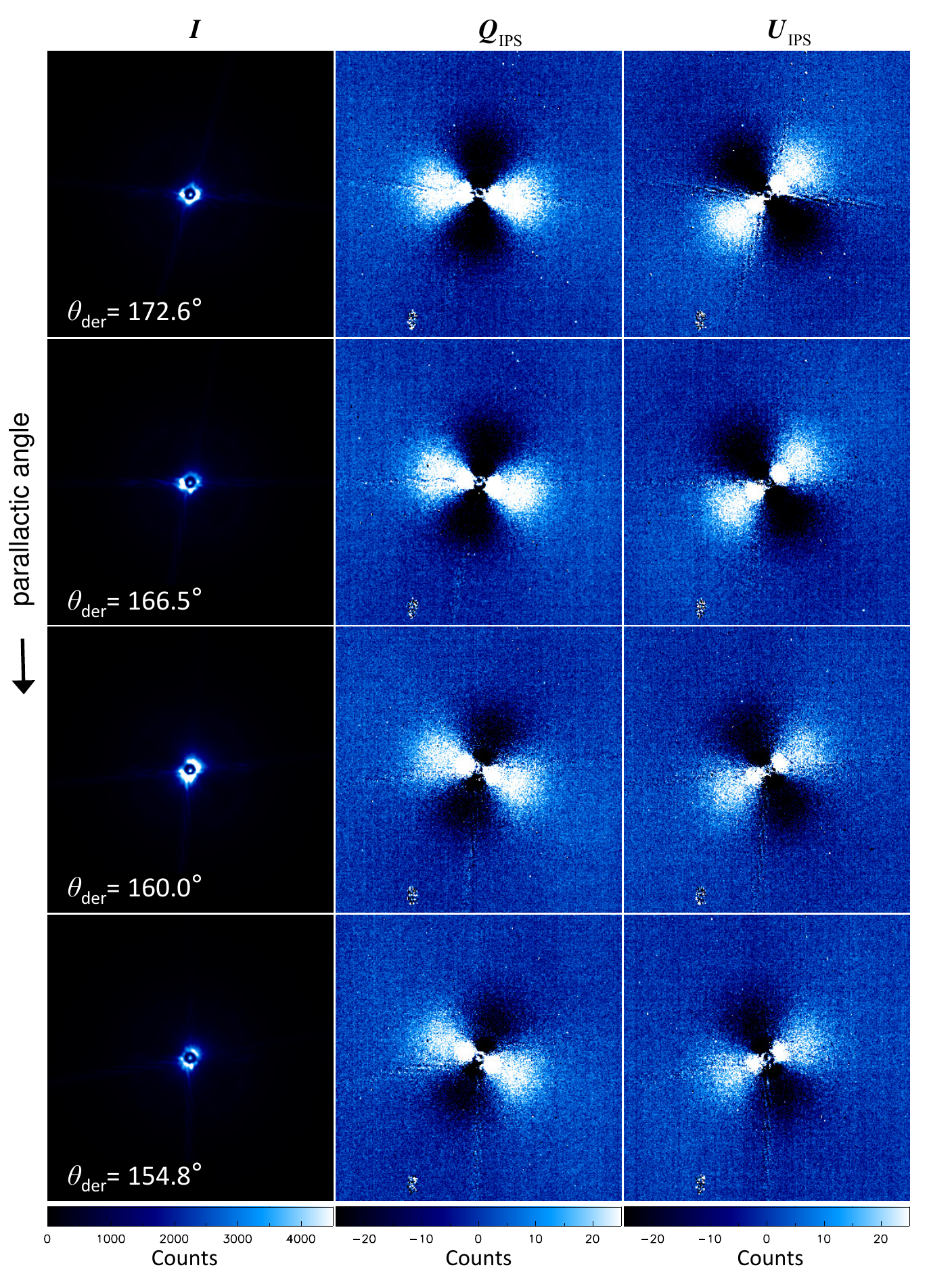}
   \caption{
	Each row shows the $I$, $Q_\mathrm{IPS}$ and $U_\mathrm{IPS}$ images (from left to right) 
	for polarimetric cycles with increasing parallactic angle from top to bottom.
	The on-sky orientation is the same for all panels: north is up, east is left. 
	However, the measured polarization angle (Eq.~\ref{eq:aolp}) is changing with derotator angle ($\theta_\mathrm{der}$), which we can see with the rotation of the butterfly pattern in clockwise direction. 
	Furthermore, the polarimetric e\ffi ciency decreases with $\theta_\mathrm{der}$ further removed from $180^\circ$, as we can see by the decreasing {\revia signal-to-noise ratio} in the $Q_\mathrm{IPS}$ and $U_\mathrm{IPS}$ images.
      \label{fig:twhbflies}}
    \end{figure*}

\subsection{Post processing: polarimetric differential imaging}
\label{sec:twhpdi}

	The single-di\ff erence images ($Q^+$, $Q^-$, $U^+$, 
	and $U^-$) are determined frame by frame for the HWP angles: $\theta_\mathrm{HWP}^s = 0^\circ$, $45^\circ$, $22.5^\circ$, 
	and $67.5^\circ$, respectively. 
	The single-di\ff erence images obtained from the four frames of each \fix le (per HWP angle) are median combined.
	$Q$ and $U$ are computed with the double-di\ff erence method 
	of Eqs.~\ref{eq:qddif} and \ref{eq:uddif}, for each polarimetric cycle (also called `HWP cycles', containing the four switch angles of HWP2:
	$\theta_\mathrm{HWP}^s = 0^\circ$, $45^\circ$, $22.5^\circ$, and $67.5^\circ$).
	Accordingly, per HWP angle the corresponding single-sum total-intensity images are created with Eq.\,\ref{eq:imeas}. 
	{\revia With these single-sum images, we create per HWP cycle the double-sum total-intensity} $I_Q$ and $I_U$ images with Eqs.\,\ref{eq:iqddif} and \ref{eq:iuddif}, respectively.

	A residual of the read-out columns (see feature {c} in Fig.\,\ref{fig:flat}) remains visible in the 
	double-di\ff erence images.
	Similar to how \citet{Avenhaus:2014} removed noise across detector rows from NACO images, 
	we remove these artifacts from the double-di\ff erence images by taking the median 
	over the top and bottom 20 pixels (to avoid including signal from the star) on the image per individual pixel column 
	(not the 64 pixel wide read-out column), and subtract this median value from the entire pixel column.

	We perform a \fix rst-order correction {\revia for \textit{IP}} created upstream from HWP2 
	(i.e. by the telescope and M4)
	on the $Q$ and $U$ images of each polarimetric cycle. 
	This correction method \citep[as described by][]{2011A&A...531A.102C} is based on the assumption that the direct stellar light is unpolarized.
	We take the median of the $Q/I$ signal over an annulus centered around the star 
	(excluding the coronagraph mask) to obtain the scalar $c_Q$ 
	(likewise, we determine $c_U$ with $U/I$),
	multiply this scalar with $I$,
	and subtract this from the $Q$ image. 
	Hence, the {\revia \textit{IP}}-{\revia subtracted} linear stokes components are:
	\begin{eqnarray}
		Q_{\mathrm{IPS}} &=& Q - I_Q \cdot c_Q , \label{eq:ipcq} \\				
		U_{\mathrm{IPS}} &=& U - I_U \cdot c_U . \label{eq:ipcu}
	\end{eqnarray}
	{\revia The size and location of the annulus over which to measure $c_Q$ and $c_U$ can be adjusted to suit a particular dataset. 
	Ideally, the annulus should lie in a region 
that should only contain non-scattered starlight, with high signal in the $I_Q$ and $I_U$ images.
Therefore, the size and location of the annulus depends on the brightness of the central star and the size and shape of the circumstellar material that has been observed. 

We now use the possible user-specific derotator o\ff set angle that can be found from the FITS header keyword INS4.DROT2.POSANG, together with the true north correction of $-1.7^\circ$ \citep{maire:2018} to apply a software derotation in order to align all $I_{Q/U}$ and $Q/U_{\mathrm{IPS}}$ with north up and east left on the detector. 
}
	
\subsection{Azimuthal Stokes parameters}
\label{sec:polstokes}
	
	To create the \fix nal polarization image we have two choices. 
	In the \fix rst, most straight-forward method we compute the polarized intensity $PI_\mathrm{L}$ according to Eq.~\ref{eq:polint} for each HWP cycle and median combine these to create a \fix nal (less noisy) $PI_\mathrm{L}$ image. 
	The problem with this method is that the squares taken in Eq.~\ref{eq:polint} boost the noise in each image.
	For example, artifacts seen as a bright positive or negative 
	feature detected at a point in the $Q_{\mathrm{IPS}}$ 
	image where the signal should be ${\sim}\,0$,
	(on the diagonal `null' lines separating the positive from the negative signal in the $Q_{\mathrm{IPS}}$ images) 
	or a strong positive signal in a region where $Q_{\mathrm{IPS}}$ ought to be negative, 
	will be indistinguishable from true disk signal in the $PI_\mathrm{L}$ image. 
	This is actually a general problem we encounter when computing $PI_\mathrm{L}$, 
	but even more so when we are dealing with images of short integration times, 
	such as resulting from individual HWP cycles.

	In stead, we have used a second option to combine the HWP cycles and create the cleanest image by computing the azimuthal Stokes parameters \citep{Schmid:2006A&A}: 
	\begin{eqnarray}
	Q_\phi &=& - Q_{\mathrm{IPS}}  \cos{(2\phi)} - U_{\mathrm{IPS}}  \sin{(2\phi)}, 
		\label{eq:qphi}\\	
	U_\phi &=& + Q_{\mathrm{IPS}}  \sin{(2\phi)} \, - U_{\mathrm{IPS}}  \cos{(2\phi)},
	\label{eq:uphi}			
\end{eqnarray}
	where $\phi$ describes the azimuth angle, which can be computed for each pixel 
	(or $x,y$ coordinate) as
	\begin{equation}	
		\phi = \arctan{\left(\frac{x_\mathrm{star} - x}{y - y_\mathrm{star}}\right)} + \phi_0.
		\label{eq:azim} 	
	\end{equation}
	{\revia  The $x$ and $y$ positions of the central star in the image are described by $x_\mathrm{star}$ and $y_\mathrm{star}$, respectively. 
We can use $\phi_0$ to give the azimuth angle an o\ff set if the measured polarization angle is not aligned azimuthally. 
$\phi_0$ is therefore referred to as the \textit{polarization angle o\ff set}.}

	{\revia Contrary to Eqs.\,\ref{eq:qphi} and \ref{eq:uphi}, \citet{Schmid:2006A&A} use the notation $Q_\mathrm{r}$ and $U_\mathrm{r}$, which have \fl ipped signs compared to $Q_\phi$ and $U_\phi$, respectively.
\citeauthor{Schmid:2006A&A} have chosen their conventions to describe scattered light observations of the planets Uranus and Neptune, which is oriented in radial direction relative to the center of the planet.
However, in protoplanery disks, we expect scattered light to produce predominantly azimuthally oriented polarization, which has motivated our choice of signs in Eqs.\,\ref{eq:qphi} and \ref{eq:uphi}. 
}
	Polarization oriented in azimuthal direction (with respect to the position of the star)
	will be measured as a positive $Q_\phi$ signal; radial polarization will show up as a negative $Q_\phi$;
	while polarization angles oriented at $\pm 45^\circ$ with respect to azimuthal will result in $\pm U_\phi$ signal.
	Disks that have a high inclination or where multiple scattering is expected to produce a signi\fix cant part of the scattered light 
	{\revia can contain} a signi\fix cant signal in $U_\phi$ \citep{Canovas:2015}.
	{\revia However,} for low-inclination disks 
	we can expect all scattering polarization to be in azimuthal direction. 
	{\revia This} means that $Q_\phi$ 
	will de facto show us $PI_\mathrm{L}$, with the bene\fix t that we do not square the noise, 
	resulting in cleaner images.
	The $U_\phi$ image should ideally show no signal at all in this case, which makes it a suitable metric for the 
	quality of our reduction.
			
\section{Instrument-con\fix guration dependence in polarimetric e\ffi ciency and polarization angle}
\label{sec:tempvar}
\subsection{Polarimetric observations of TW\,Hydrae}
\label{sec:twh}

	We have observed TW\,Hydrae during the night of March 31, 2015, with IRDIS/DPI.
	This data was recorded before we became aware of the most severe instrumental polarization e\ff ects for SPHERE/IRDIS. Therefore, these data have been recorded without taking recommendations (Sect.\,\ref{sec:recommend}) into account that optimize the polarimetric e\ffi ciency of DPI observations.
	Furthermore, the {\revia near} face-on orientation of this disk {\revia \citep[inclination $= 7^\circ$,][]{Qi:2008}, allows us to assume azimuthal polarization after scattering, which makes} this object an ideal test case to illustrate how instrumental polarization e\ff ects alter the incident {\revia polarized} signal.

	The data have been recorded in $H$ band (see Table~\ref{tab:filters}), using \fix eld-stabilized mode, while an apodized pupil Lyot 
	coronagraph \citep{Carbillet:2011, Guerri:2011} with a focal plane mask with radius of 93\,mas was used.
	{\revia We have performed the observations using} a detector integration time (DIT) of 16\,s per frame, four frames per \fix le, 
	during 25 polarimetric cycles.
	This adds up to a total exposure time of 106.7\,min.

	{\revia After creating the double-difference images} we removed \fix ve HWP cycles with bad seeing and/or AO corrections.
	{\revia Therefore, the final dataset we have used for this analysis contains} 20 sets of $Q$, $U$, $I_Q$ and $I_U$ images.

	Figure~\ref{fig:twhbflies} shows the $I = (I_Q + I_U)/2$, $Q_{\mathrm{IPS}}$ 
	and $U_{\mathrm{IPS}}$ images 
	for four polarimetric cycles observed with increasing parallactic angle for each subsequent panel row.
	While each $Q_{\mathrm{IPS}}$ and $U_{\mathrm{IPS}}$ 
	panel displays the typical `butterfly' signal 
	of an approximately face-on and axi-symmetric disk, strong variations occur between the polarimetric cycles:
	the butterflies appear to rotate in clockwise direction and the signal in the $Q_{\mathrm{IPS}}$ 
	and $U_{\mathrm{IPS}}$ images decreases with increasing {\revia parallactic angle}.
	Since the observations were taken in \fix eld tracking mode, the image of the disk itself does not rotate on the detector,
	rather the $AoLP$ (Eq.~\ref{eq:aolp}) 
	is changing between the polarimetric cycles.
	We do not expect either the incident (`true') Degree or Angle of Linear Polarization to vary with {\revia parallactic angle}.
	Notice that $I$ remains roughly constant in Fig.\,\ref{fig:twhbflies}, 
	{\revia thus a decrease in measured $PI_\mathrm{L}$ can only be explained by a decrease in $P_\mathrm{L}$}.
	Therefore, the changes in $P_\mathrm{L}$ and $AoLP$
must be caused by instrumental polarization e\ff ects {\revia that depend on the speci\fix c telescope and instrument con\fix guration, which varies with the parallactic and altitude angle of the observed star.}

\begin{figure*}[!h]
   \centering
   \includegraphics[height=7.5 cm, trim = 0 -2 7 10]{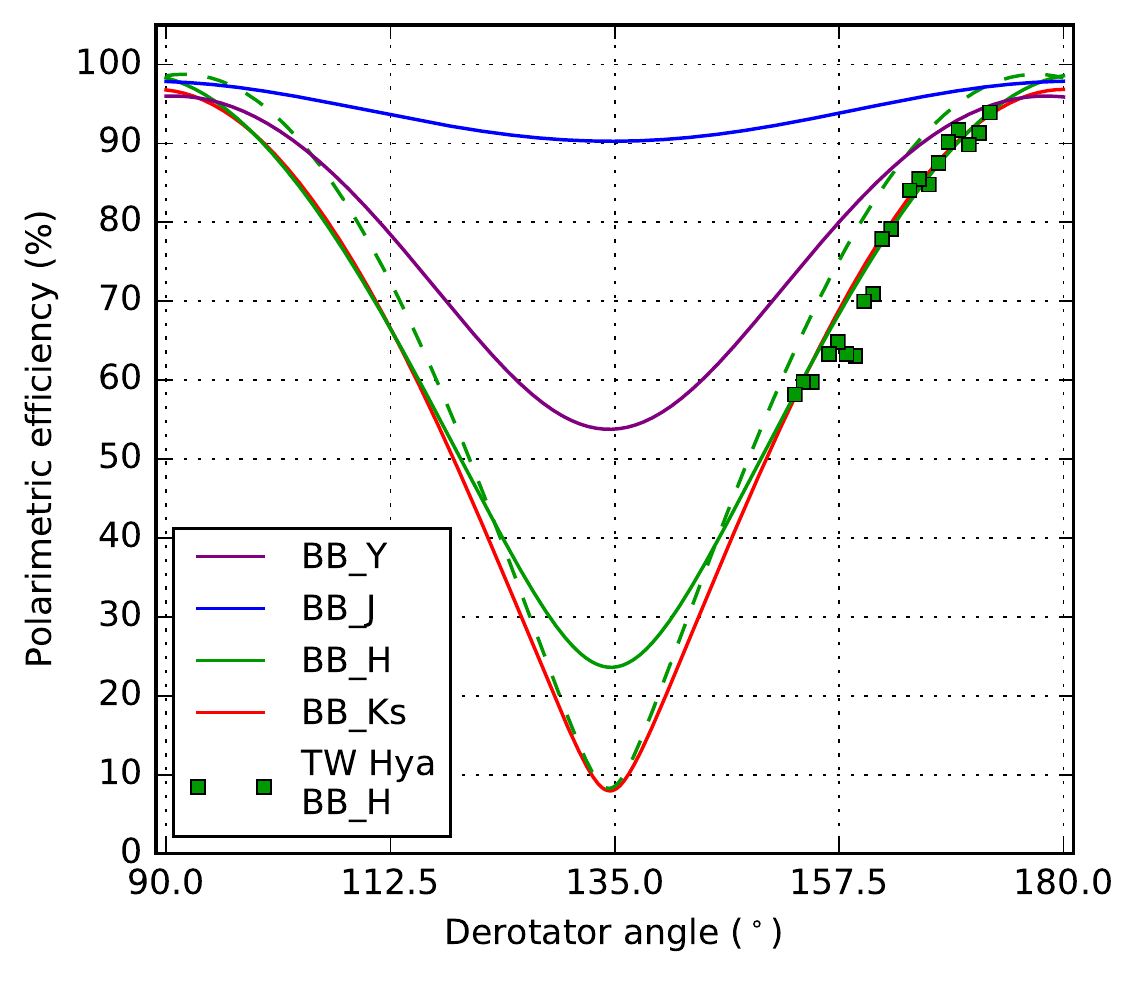}%
   \includegraphics[height=7.5 cm, trim = -3 0 0 15]{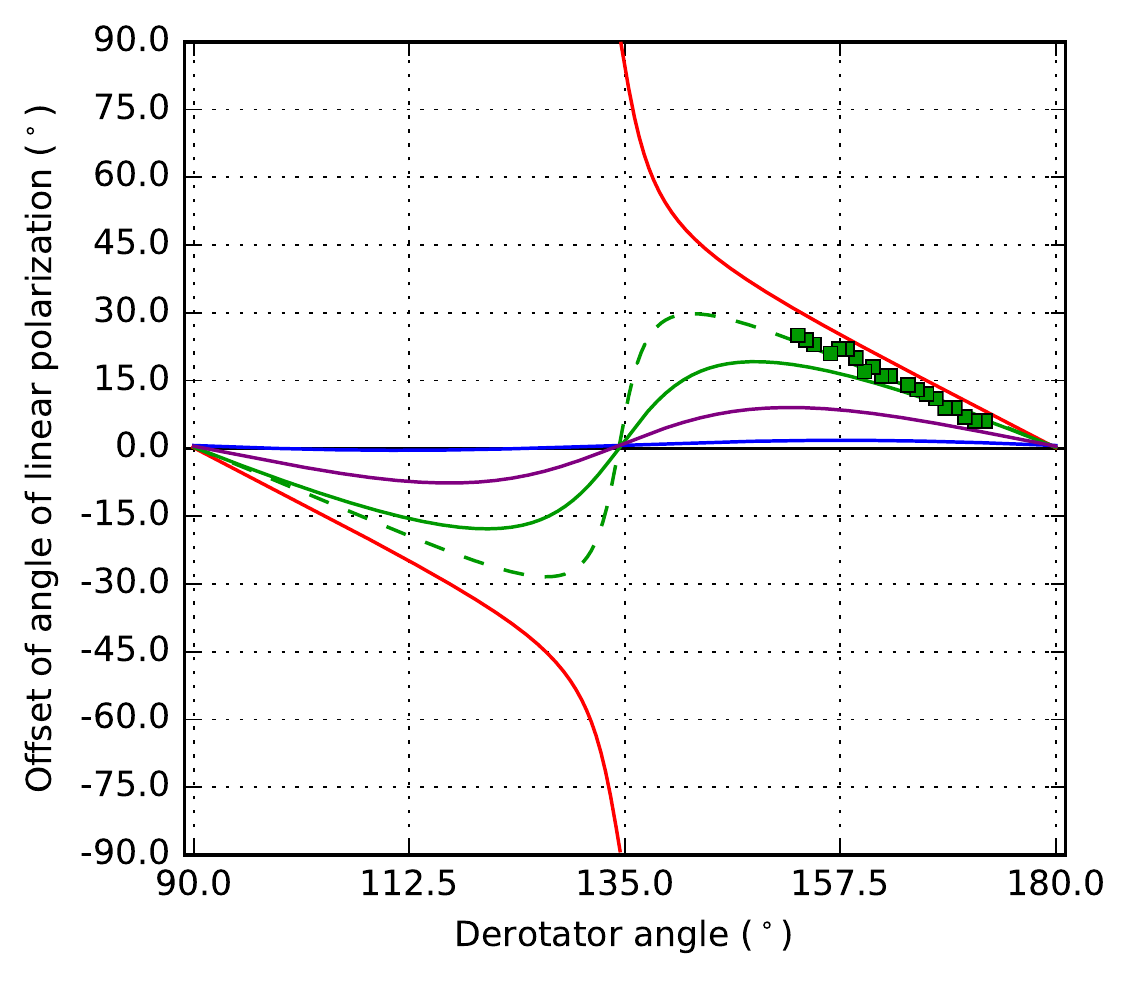}%
   \caption{
\textbf{Left: }
   	Polarimetric e\ffi ciency as a function of derotator angle for all broadband \fix lters
   	listed in Table~\ref{tab:filters}.
   	The green squares show the $H$-band polarimetric e\ffi ciencies measured for TW\,Hydrae scaled to the model curves 
	(Sect.\,\ref{sec:expltwh}).
	The green dashed curve shows the model polarimetric e\ffi ciency for the range 
	$0^\circ \leq \theta_\mathrm{der} \leq 90^\circ$ in $H$ band. 
	This \fix lter shows the strongest asymmetry around $\theta_\mathrm{der} = 90^\circ$, which is caused by the 
	non-ideal retardance of HWP2.
\textbf{Right: } 
	The model polarization angle o\ff set plotted against $\theta_\mathrm{der}$ for the same \fix lters {\revia (solid and dashed lines; as before, the dashed line covers the range $0^\circ \leq \theta_\mathrm{der} \leq 90^\circ$)}.
   	In the ideal case, the polarization angle o\ff set would remain $0^\circ$. 
   	However, there is a clear dependency on $\theta_\mathrm{der}$ for $H$ and $K_s$ band, 
   	and to a lesser extend for $Y$ band, while the $J$ band remains close to ideal.
	The polarization angle o\ff sets measured as $\phi_0$ for TW\,Hydrae {\revia are plotted as} green squares.
	The deviation of the on-sky data from the model curves is caused by the crosstalk contribution of other optical components.
      \label{fig:poleff}}
    \end{figure*}

	\subsection{Instrumental polarization e\ff ects}
	\label{sec:calib}

	The calibrations of instrumental polarization e\ff ects 
	and the analysis towards a complete Mueller matrix model of the instrument 
	are described in detail in Paper\,II.
	In this section we 
	will briefly summarize {\revia how we have derived the model from calibration measurements, and describe} the instrumental polarization e\ff ect of each \textit{set of optical components}.
{\revia Here we consider optical components to form a `set' when they} share a \fix xed {reference frame}, i.e.~a common rotation of the set of components.
	{\revia For the last set of components} (CPI + IRDIS components downstream of the derotator) 
	{\revia we only \fix t the diattenuation, since crosstalk is absent because all re\fl ections are aligned with the anlyzers}. 

{\revia
	In Sect.\,\ref{sec:expltwh} we use this polarimetric instrument model to explain variations detected in the degree and angle of linear polarization in the data of TW\,Hydrae.
}
	Based on the polarimetric instrument model we have devised a correction method in Paper\,II.
	In Sect.\,\ref{sec:comparemeth} of this paper we apply this correction method to retrieve the true incident polarization for the observations of TW\,Hydrae.

	\subsubsection{From calibrations to polarimetric instrument model}
	\label{sec:intromodel}

	{\revia
	In Paper\,II, we have used the internal light source with a calibration polarizer to create $100\%$ polarized light, and measured the linear Stokes parameters
for a wide range of instrumental con\fix gurations. 
	We then de\fix ned the measured degree of linear polarization for the $100\%$ polarized incident light to be equal to the polarimetric e\ffi ciency for this con\fix guration. 
	For each set of optical components including and downstream of the HWP, we have \fix tted the measured Stokes parameters
to the wavelength-dependent retardance of this set of components.
	Similarly, we have calibrated the diattenuation of these sets of optical components with the internal light source, this time without the inclusion of a calibration polarizer to insert nearly unpolarized incident light. 
		
	We have calibrated the diattenuation of the telescope and M4 (both located upstream from the HWP) by observing unpolarized stars at various altitude angles of the telescope. 
	The retardances of these optical components have been determined analytically using the Fresnel equations and literature values for the complex refractive index of the coating material. 
	With the diattenuation and retardance we have computed the wavelength-dependent $4\,{\times}\,4$ Mueller matrices for each set of optical components that share a reference frame. The combination of the Mueller matrices for all optical components forms our polarimetric instrument model.
}

\subsubsection{The derotator and HWP2}
\label{sec:derot}

	The left-hand panel of Figure~\ref{fig:poleff} shows the polarimetric e\ffi ciency curves against derotator angle ($\theta_\mathrm{der}$) for the four broadband \fix lters 
	of IRDIS.
	The solid lines show the {\revia polarimetric e\ffi ciency derived from the} Mueller matrix model for the optical system 
	{\revia for $90^\circ \leq \theta_\mathrm{der} \leq 180^\circ$, while the green dashed line shows this for $0^\circ \leq \theta_\mathrm{der} \leq 90^\circ$ in $H$-band only. 
	These green solid and dashed curves clearly do not overlap, and the same is true for the other filters (not shown)}.	
This asymmetry across $\theta_\mathrm{der} = 90^\circ$ in the polarimetric e\ff iciency curve is caused by a non-ideal behavior of HWP2, i.e. 
	the retardance $\neq \lambda/2$ 
(see Paper\,II).
	A dramatic decrease in polarimetric e\ffi ciency is seen for 
	$\theta_\mathrm{der}\,{\approx}\,45^\circ$ and $135^\circ$ in the
	$H$ and $K_s$-band \fix lters, 
	while $Y$-band and especially the $J$-band \fix lters show a much better polarimetric e\ffi ciency curve.

	The right-hand panel of Figure~\ref{fig:poleff} shows how the {\revia polarization angle o\ff sets}
oscillate around $0^\circ$ (which would be the value in the ideal case of no crosstalk) {\revia for the model (solid and dashed lines) and the data (green squares)}.
	While this oscillation is only marginally visible for $J$ band, with a maximum deviation from ideal $< 4^\circ$;
	it is $< 11^\circ$ in $Y$, and can reach up to $34^\circ$ in $H$.
	For $K_s$ band, {\revia the polarization angle o\ff set} does not even return to its equilibrium and continues rotating beyond 
	$\pm 90^\circ$ (where a rotation of $+90^\circ$ is indistinguishable from $-90^\circ$).
	
	The {\revia strong} dependence of {\revia the polarization angle o\ff set} and polarimetric e\ffi ciency on $\theta_\mathrm{der}$ in $H$ and $K_s$ band are predominantly caused by crosstalk induced by the retardance of the derotator, 
	which is close to that of a quarter-wave ($\lambda/4$) plate at these wavelengths. 
	With these retardances, the derotator causes a strong linear to circular polarization crosstalk. 
	This crosstalk means that when we use the wrong observing strategy we can lose up to $95\%$ 
	of incident linearly polarized signal, which is ultimately the information carier we aim to measure.
	{\revia 
	Because other optical components (e.g., HWP2) contribute to this loss of polarization signal as well, we list the polarimetric e\ffi ciencies for the least favorable instrumental con\fix guration in Table\,\ref{tab:poleff}}.

\begin{table}[!h]
\centering
\caption{Lowest polarimetric e\ffi ciencies reached at the least favorable instrumental set-up. 
The values are determined with the polarimetric instrument model (Paper\,II) for the broad-band filters described in Table\,\ref{tab:filters}.}
 \centering 
\begin{tabular}{c c c c} 
\hline\hline
\T	    $Y$ ($\%$) & $J$ ($\%$) & $H$ ($\%$) & ${K_s}$ ($\%$)   \\ \hline 
\T 54 & 89 & 5 & 7 \\ \hline
\end{tabular}  
\label{tab:poleff}
\end{table}

	 \subsubsection{The telescope and SPHERE's \fix rst mirror}
\begin{table}[!h]
\centering
\caption{{\revia \textit{IP}} (in percent of the total intensity) induced by the telescope mirrors and M4 before and after recoating of the M1 and M3.
The values are shown at optimal, nearly crossed con\fix guration of the re\fl ection planes of M3 and M4 (at $a = 87^\circ$, with $a$ the altitude angle of the Unit Telescope), 
and worst, close to aligned re\fl ection planes ($a = 30^\circ$: the lowest altitude at which the ADC can correct for atmospheric dispersion) for the IRDIS' broadband \fix lters.
		}
 \centering 
\begin{tabular}{l c c c c c} 
\hline\hline
\T	Date     & $a$ ($^\circ$)  & $Y$ ($\%$) & $J$ ($\%$) & $H$ ($\%$) & ${K_s}$ ($\%$)   \\ \hline 
\T 	Before   		&  87 	& 0.58	& 0.42	& 0.33   	&  0.29 \\ 
 	16-04-2017	&  30   	& 3.5	& 2.5	& 1.9      	&  1.5 \\ \hline
\T	After     		&  87   	& 0.18	& 0.12	& 0.07	&  0.06 \\ 
\B  	16-04-2017  	&  30  	& 3.0	& 2.1		& 1.5	     	&  1.3 \\
 	\hline 
\end{tabular}  
\label{tab:m3m4}
\end{table}

	
	On April 16, 2017, UT3's M1 and M3 have been recoated, resulting in a more e\ff ective cancellation of IP when M3 and M4 are in crossed con\fix guration (when looking at or close to zenith).
	Therefore, we present the lowest and highest {\revia \textit{IP}} values for each broadband \fix lter as measured before and after recoating in Table\,\ref{tab:m3m4}.

	\subsection{Explaining TW\,Hydrae data with the instrument model}
	\label{sec:expltwh}

	 \begin{figure}[h]
   \centering
   	\includegraphics[width=0.49\textwidth, trim = -3 0 0 -2]{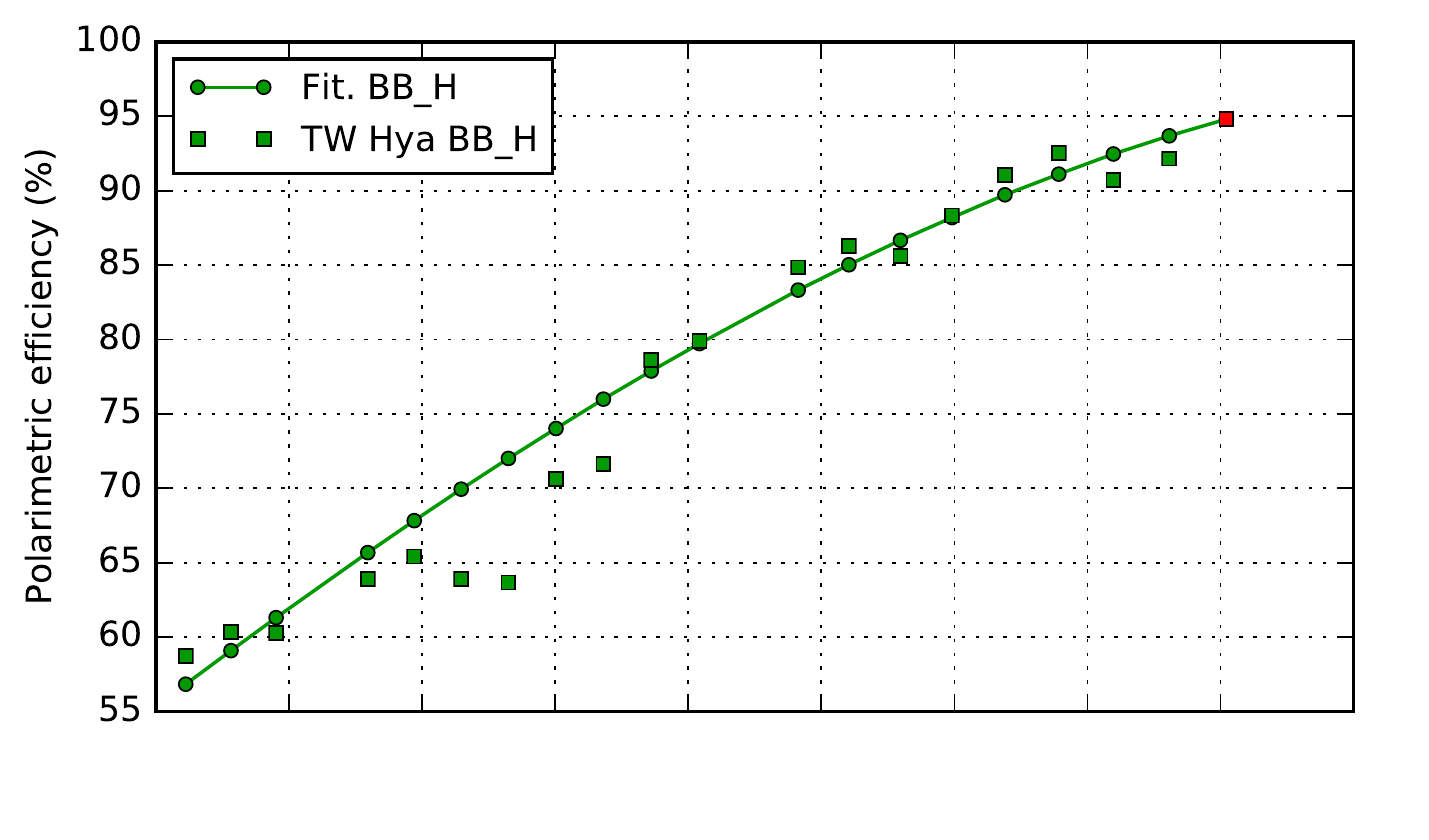}
      	\includegraphics[width=0.49\textwidth, trim = 0 0 0 35]{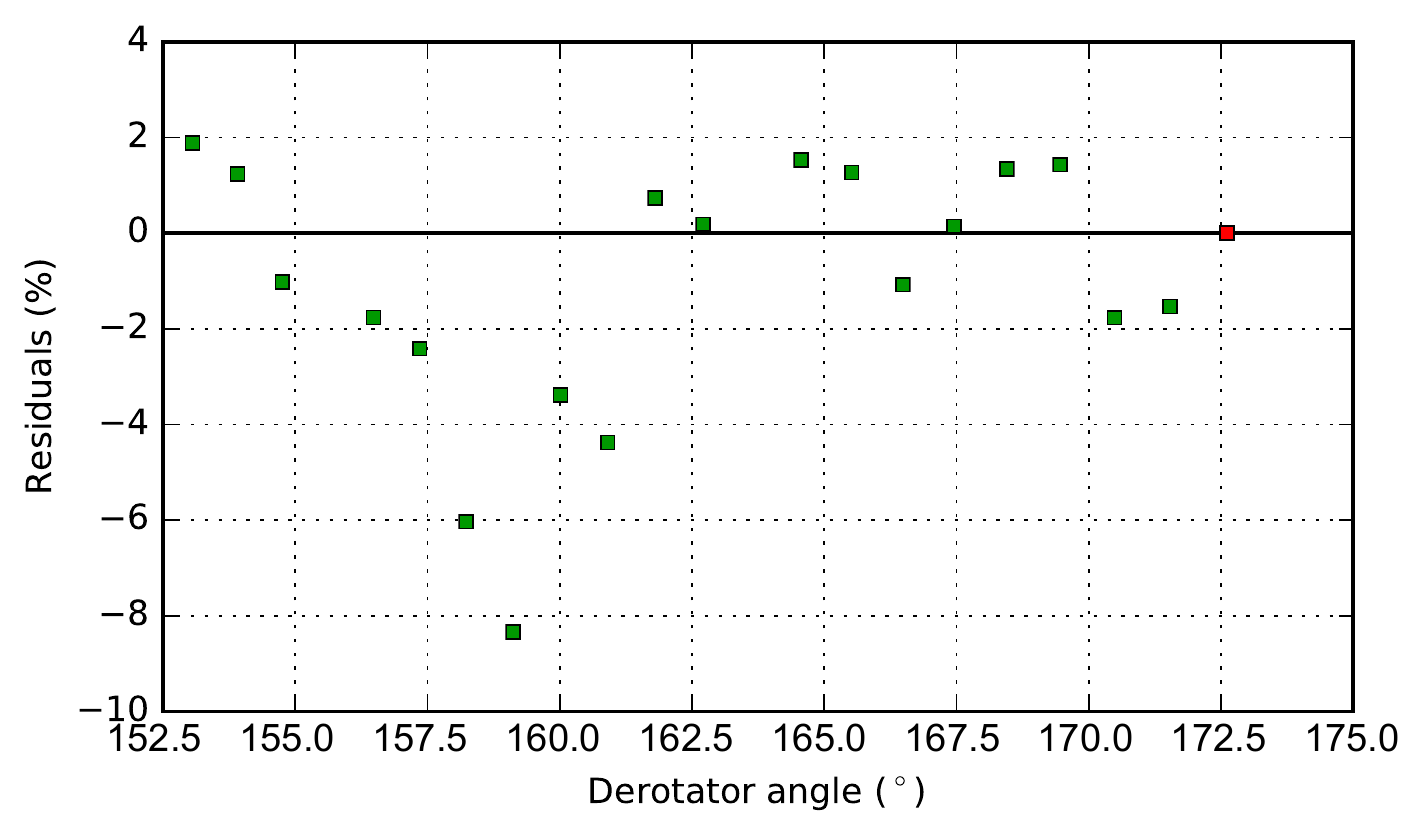}
   \caption{
   \textbf{Top:} Polarimetric e\ffi ciency modeled (solid line) for the same derotator and HWP2 angles as used during the observation of TW\,Hydrae (squares). 
   Note that the polarimetric e\ffi ciencies measured for the HWP cycles of TW\,Hydrae are only determined relative to the other HWP cycles.
   We therefore scaled all data points such that the first cycle (red square, with the highest polarimetric e\ffi ciency) matches the value of the model.
   \textbf{Bottom:} Residuals between the model and the polarimetric e\ffi ciencies obtained for TW\,Hydrae.
      \label{fig:dolptwh}}
    \end{figure}
    
 	 \begin{figure}[!h]
   \centering
   	\includegraphics[width=0.49\textwidth, trim = 0 0 0 0]{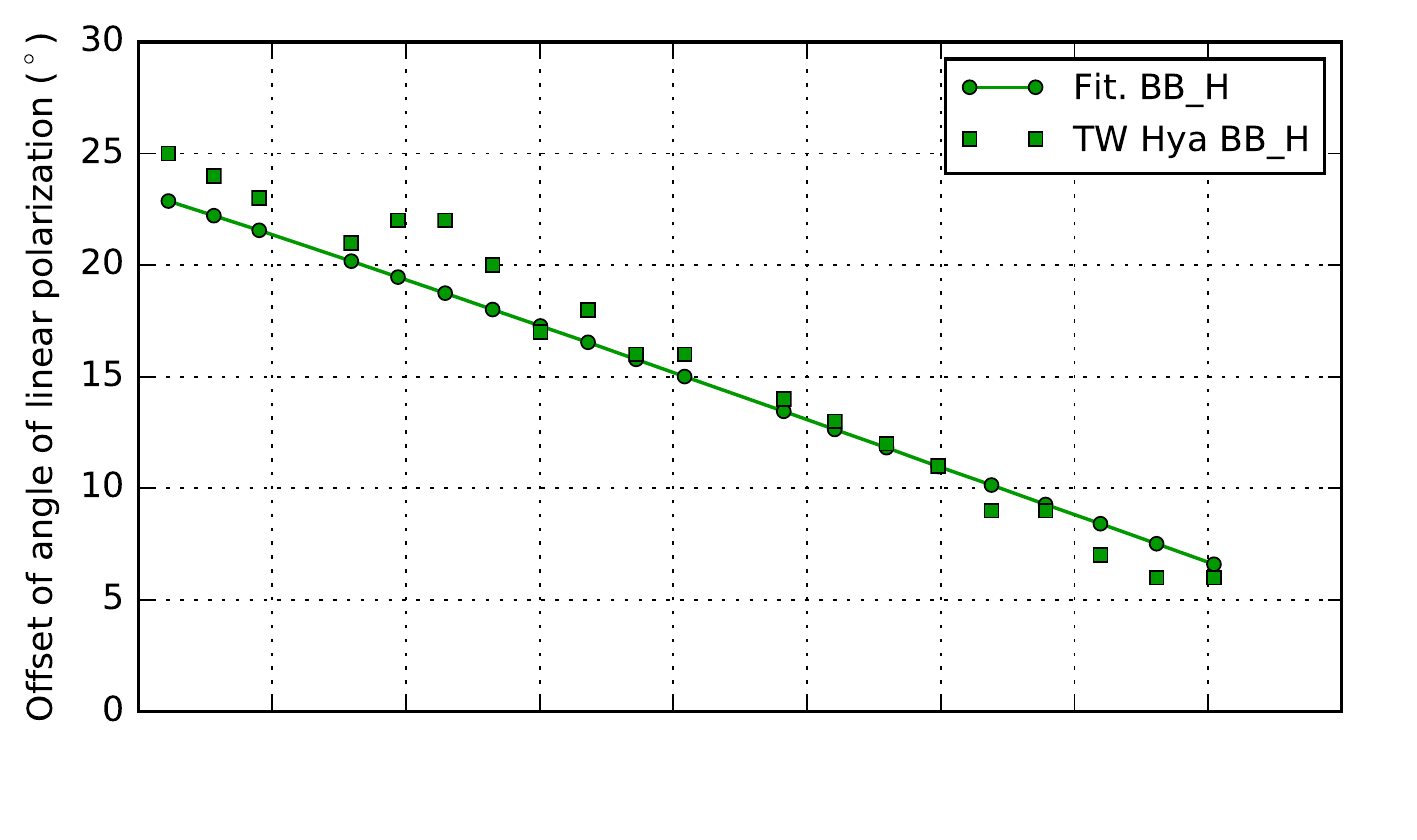}
      	\includegraphics[width=0.49\textwidth, trim = 0 0 0 36]{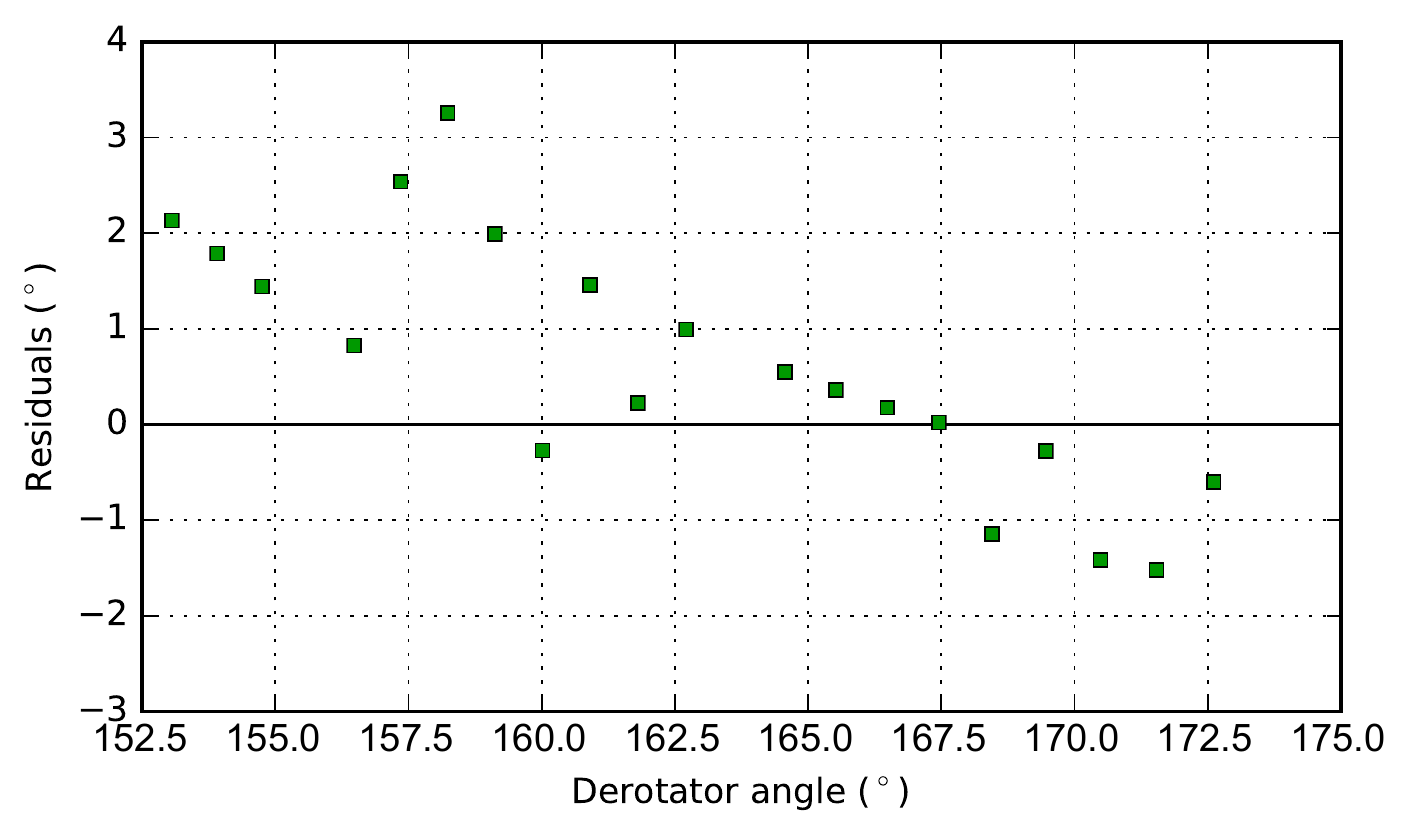}
   \caption{
   \textbf{Top:} {\revia Expected polarization} angle o\ff sets of the same model (solid line) as used in Fig.\,\ref{fig:dolptwh};
   and the polarization angle o\ff set angles ($\phi_0$ of Eq.~\ref{eq:azim}) of TW\,Hydrae (squares) with respect to azimuthal polarization.
   \textbf{Bottom:} Residuals between the model {\revia polarization} angle o\ff sets and $\phi_0$ retrieved for TW\,Hydrae.
      \label{fig:aolptwh}}
    \end{figure}

	During the observation of the 20 HWP cycles, 
	the derotator has rotated from \derot $\,= 173.2^\circ$
	to \derot $\,= 152.6^\circ$ ($\Delta$\derot $\,= -20.6^\circ$). 
	To account for the variation of the measured $AoLP$ between the HWP cycles, 
	we have determined the correct value for the polarization angle o\ff set $\phi_0$ for each cycle separately,
	based on the assumption that the polarization is oriented in azimuthal direction, and therefore 
	$U_\phi$ should be 0.
	We have achieved this by {\revia computing the sum over the} absolute signals {\revia measured for the pixels in} a centered annulus in the $U_\phi$ image for a range of 
	$\phi_0$ values: $c_{U_\phi}(\phi_0)$. 
	We have selected the $\phi_0$ value which yielded the lowest $c_{U_\phi}(\phi_0)$.
	We have derived the relative polarimetric e\ffi ciency by measuring the absolute signal over an annulus in the 
	$Q_\phi$ image for each cycle, and dividing these values by that of the highest (coincidentally the \fix rst) HWP cycle.
	During the observing sequence of TW\,Hydrae the polarimetric e\ffi ciency has decreased with ${\sim}\,40\%$.

	The green squares in Fig.\,\ref{fig:poleff} represent the relative polarimetric e\ffi ciency (left panel) and the polarization angle o\ff set (right panel) for the TW\,Hydrae
 {\revia measurements. For this dataset we know neither the incident nor measured} $P_\mathrm{L}$ ({\revia instead, we measure $Q_\phi \approx PI_\mathrm{L}$ and do not know $I$ of the disk}) 
{\revia to determine the absolute value of the polarimetric e\ffi ciency.
However, during the observations, \qp is expected to be linearly proportional to the polarimetric e\ffi ciency.
Therefore, we have measured for each HWP cycle the mean \qp in a \fix xed annulus around the star and scale the images such that the highest mean value} (from the \fix rst HWP cycle) matches the model's polarimetric e\ffi ciency in $H$.

	Although the polarimetric e\ffi ciency is rather well 
	explained with the $H$-band model curve (green solid line), 
	the polarization angle o\ff set 
	deviates from the model.
	{\revia The models shown in Fig.\,\ref{fig:poleff} are created for the simplest con\fix guration, where all instrument settings have been kept} constant except \derot,
while during the {\revia on-sky} observations many optical components have changed position due to their {\revia \fix eld-tracking laws, such as HWP2 and the telescope altitude angle}.

\begin{figure*}[!h]
   \centering
   \includegraphics[width=0.7\textwidth, trim = 0 20 0 0]{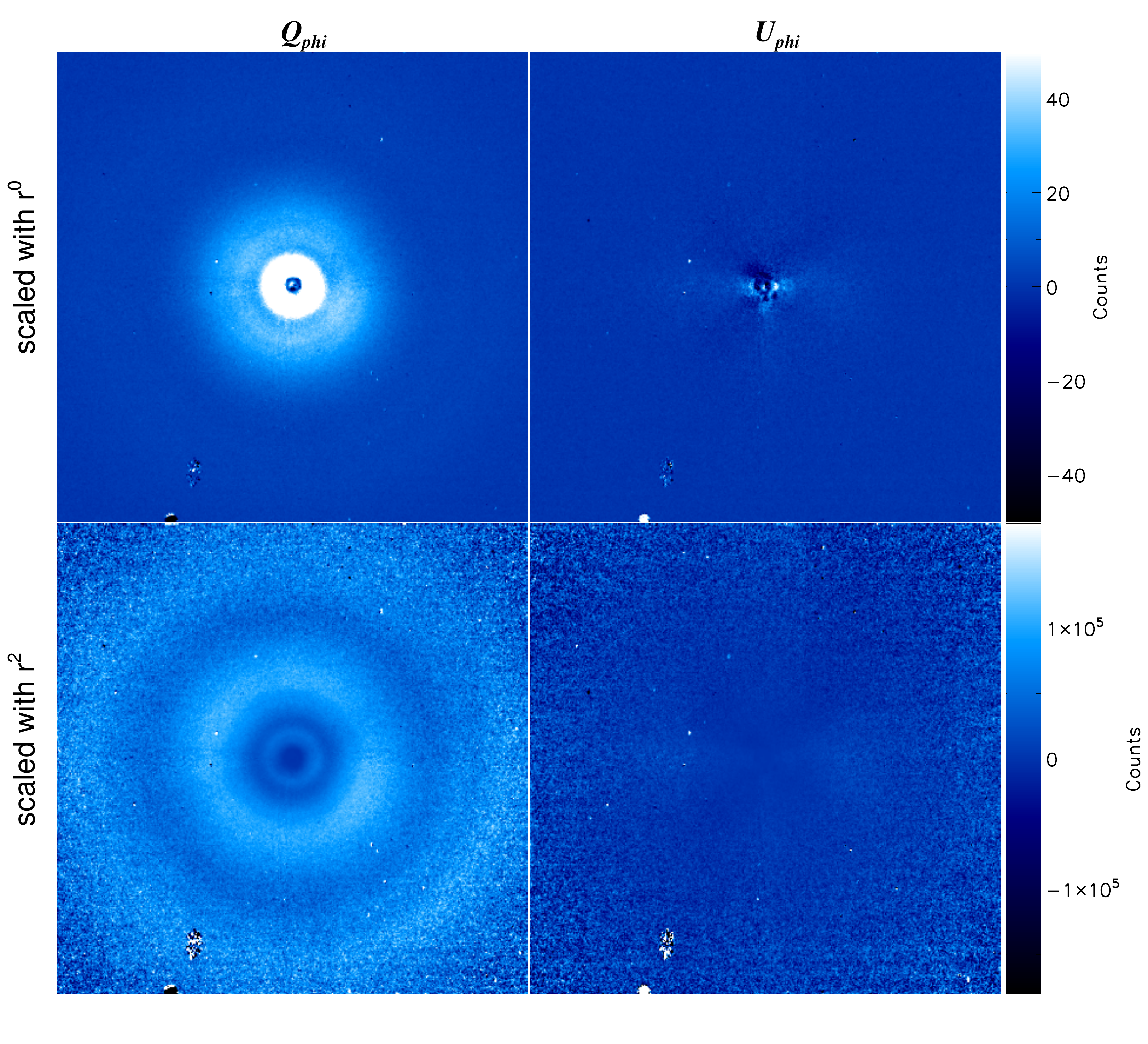}
   \caption{
	Final $Q_\phi$ (left) and $U_\phi$ (right) images of TW Hydrae. 
	Both $Q_\phi$ and $U_\phi$ are displayed
	 with identical linear scale, and either unscaled (top) or scaled with $r^2$ (bottom) 
	to compensate for the decrease of stellar \fl ux with distance.
	 All four panels are shown with North up and East to the left for the same \fix eld of view of 
	 $4.9''\,{\times}\,4.9''$ or up to a separation of 
	 $r = 2.45''$ from the star in both RA ($-x$-axis) and Dec ($+y$-axis).
      \label{fig:qphiuphi}}
    \end{figure*}        	

	{To account for these additional changes in con\fix guration, we use} the instrument model {\revia to} compute the polarimetric e\ffi ciency and polarization angle o\ff set  
	for the {\revia same instrument con\fix guration as was} used during the observations of TW Hydrae.
	We compare the predicted polarimetric e\ffi ciencies and {\revia polarization angle} o\ff sets
	with on-sky observation polarimetric e\ffi ciencies in Fig.\,\ref{fig:dolptwh} 
	and {\revia polarization} angle o\ff sets in Fig.\,\ref{fig:aolptwh}.
	
	The model predictions are very succesful at explaining the changes in polarimetric e\ffi ciency and polarization angle.
	 We see some clear outliers in both Figs.\,\ref{fig:dolptwh} and \ref{fig:aolptwh} at $\theta_\mathrm{der} \approx 139^\circ$,
	 which are most likely caused by a poor \fix t of $\phi_0$ for these HWP cycles.
	This shows that our comparison is limited by the accuracy of the polarimetric e\ffi ciency measurement in the data rather than that of the model.

\section{Comparison between data reduction with $U_\phi$ minimization and model-based correction}
\label{sec:comparemeth}

After Sect.\,\ref{sec:twh} we paused our post-processing of the data to analyze the instrumental 
	polarization e\ff ects that cause the detected variations {\revia in $P_\mathrm{L}$ and $AoLP$ for TW\,Hydrae} in Sect.\,\ref{sec:calib} and \ref{sec:expltwh}.
	We have succesfully implemented these lessons to create a detailed polarimetric instrument model 
	and a model-based correction method for the instrumental polarization e\ff ects (see Paper\,II).
	In this section we will compare post-processing based on this correction method with the best post-processing we have performed without using the correction method, 
{\revia where we have corrected for residual $IP$ empirically by minimizing signal in $U_\phi$.
We will continue with the empiric correction method from where we stopped in Sect.\,\ref{sec:twh}.}

\subsection{Refining the reduction by minimizing $U_\phi$}															\label{sec:twresults}

	We determine $\phi_0$ as described in Sect.\,\ref{sec:expltwh}.
	Then we improve our centering by shifting the $Q_{\mathrm{IPS}}$
	and $U_{\mathrm{IPS}}$ images with a range of $x$ and $y$ steps
	to \fix nd the minimum $c_{U_\phi}(x,y)$ value.
	Because the improved centering will affect the minimization process with which we found $\phi_0$,
	we repeat the minimization of $c_{U_\phi}(\phi_0)$ on the centered data, and	
	\fix nd $\phi_0$ with increasing values between
	$6^\circ \leq \phi_0 \leq 25^\circ$ for the 20 HWP cycles.
	A final $U_\phi$ minimization is performed to enhance our {\revia \textit{IP}} correction:
	we \fix nd the minimum of $c_{U_\phi}$
	by searching a grid of constants $c^i_{Q}$ and $c^j_{U}$ with which we replace $c_Q$ and  $c_U$ in Eqs.\,\ref{eq:ipcq}\,\&\,\ref{eq:ipcu} to compute $Q_\mathrm{IPS}$ and $U_\mathrm{IPS}$, respectively. 
	For each point ($i,j$) in the grid, we compute $U_\phi$ with Eq.\,\ref{eq:uphi} and
	addopt the $c^i_{Q}$ and $c^j_{U}$ values that yield the smallest value of $c_{U_\phi}$.
	
	The $U_\phi$ minimizations are performed and
	$Q_{\phi}$ \& $U_{\phi}$ are computed per HWP cycle.
	At this stage we determine the polarimetric e\ffi ciency as described in Sect.\,\ref{sec:expltwh}.
	The \fix nal $Q_\phi$ \& $U_\phi$ images shown in Fig.\,\ref{fig:qphiuphi}
	are created by median combining the 20 $Q_{\phi}$ and $U_{\phi}$ images determined per cycle, respectively.

\subsection{Correcting observations with the instrument model}
\label{sec:inversemod}

	\begin{figure*}[!h]
   		\centering
   		\includegraphics[width=1\textwidth, trim = 0 0 0 0]{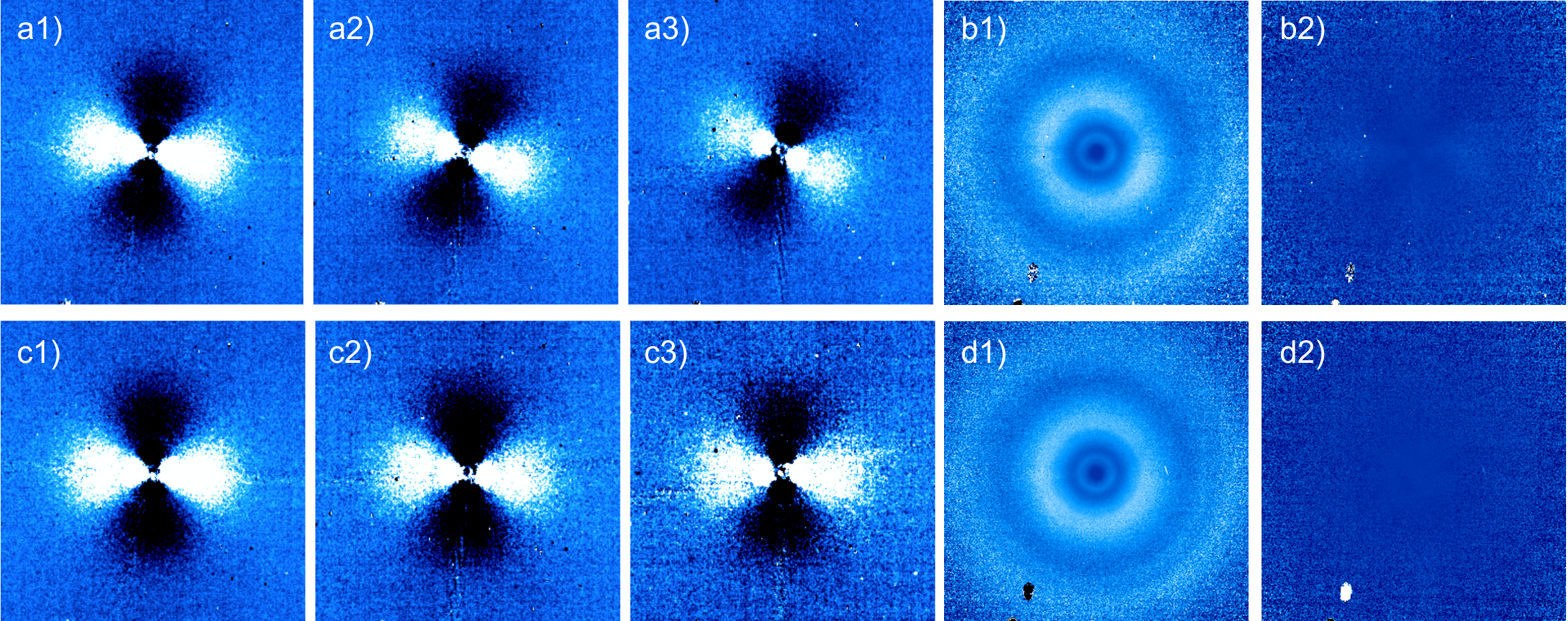}
   		\caption{
   		\textbf{a1, a2, \& a3)} $Q_\mathrm{IPS}$ images of three HWP cycles with $\theta_\mathrm{der} = 169.7^\circ, 162.0^\circ$, and $154.1^\circ$, respectively. 
		These images are created with the $U_\phi$ minimization method (Sect.\,\ref{sec:twresults}).
   		We can clearly see the buttefly patterns rotate in clockwise direction and the signal decrease from left to right.
		\textbf{b)} When we use the assumption of azimuthal polarization, we can \fix nd the correct $\phi_0$
		for each HWP cycle and compute the final $r^2$-scaled $Q_\phi$ (\textbf{b1}) and $U_\phi$ (\textbf{b2}) images (Sect.\,\ref{sec:twresults}).
		Because we have used the assumption that we know the Angle of Linear Polarization at each point, 
		we cannot claim to have determined the $AoLP$ in this image.
		\textbf{c)} When we apply the correction method to the $Q$ and 
		$U$ images, we retrieve the best approximation of the incident $Q$ and $U$ images 
		(Sect.\,\ref{sec:inversemod}).
		The displayed $Q$ images (for the same $\theta_\mathrm{der}$ as panel a) contain the ideal orientation of the butterflies and are corrected for 
		the reduced polarimetric e\ffi ciency. However, the latter does not improve the signal to noise 
		(illustrated by the increased noise in panel {c3}).
		\textbf{d1 \& d2)} The IRDAP-corrected $Q_\phi$ and $U_\phi$ image.
		Because we no longer need the assumption that we know the $AoLP$ a priori,
		we can use $Q$ and $U$ after the correction method to determine the $AoLP$ in our final reduction.
      		\label{fig:twhcor}}
    	\end{figure*}	


	In Sect.\,\ref{sec:expltwh} we show that the polarimetric instrument model can very well explain 
	the variations {\revia in $P_\mathrm{L}$ and $AoLP$} measured in the data as instrumental polarization e\ff ects.
	{\revia In Paper\,II, we have presented a highly-automated data-reduction pipeline (IRDAP) that contains a correction method based on this instrument model.
	We have reduced the TW\,Hydrae data with IRDAP, without any prior correction for instrumental polarization e\ff ects (i.e.~we did not apply Eqs.\,\ref{eq:ipcq} and \ref{eq:ipcu} or \up minimization)}.
	For illustrative purposes, we have also applied the model correction 
	to three individual 
	polarimetric cycles with $\theta_\mathrm{der} = 169.7^\circ, 162.0^\circ$, and $154.1^\circ$ of this dataset.
	Fig.\,\ref{fig:twhcor} shows the images as reduced with the method described in Sect.\,\ref{sec:twresults} and the result of the IRDAP corrections.
	The $Q_\mathrm{IPS}$ images of these three HWP cycles reduced with $U_\phi$ minimization are shown in panels a1, a2 and a3, 
	the final $Q_\phi$ and $U_\phi$ images of this reduction method are shown in panels b1 and b2, respectively.
	The $Q$ images for the same HWP cycles of panel a are shown after application of the correction method
	in panels c1, c2 and c3, and the final IRDAP-reduced $Q_\phi$  and $U_\phi$ images are shown in panels d1 and d2, respectively.

\subsection{Comparison of reduction methods}
	
	While the $Q_\mathrm{IPS}$ images of Fig.\,\ref{fig:twhcor}a show the rotation 
	of the butterfly ($\phi_0 \ne 0$)  caused by crosstalk, 
	the corrected $Q$ images of panel c are clearly oriented such that $\phi_0 \approx\,0^\circ$.
	Although the IRDAP-corrected $Q$ images display a surface brightness which is approximately the same for all 
	three images in panel c, the loss of polarization signal in the three $Q_\mathrm{IPS}$ images
	(from a1 to a3) is still visible as a decrease of the signal-to-noise ratio when comparing panel c3 with c1. 
	
	The azimuthal direction of the true polarization angle was used as an assumption in our reduction of  
	$Q_\phi$ in Sect.\,\ref{sec:twresults}. 
	Therefore, we cannot claim to have derived the angle of linear polarization in panel b1.
	However, since we do not need to assume a-priori knowledge about $AoLP$ 
	to compute the final $Q_\phi$ image of panel d1, 
	we can confidently claim to have determined the $AoLP$. 
	{\revia As a result, the $U_\phi$ image of panel d2, created with $\phi_0 =\,0^\circ$ is even cleaner (especially at small separations) than the $U_\phi$ image of panel b2}.

{\revia Although the $Q_\phi$ images produced with both methods (Fig.\,\ref{fig:twhcor} b1 \& d1) show to a large degree the same disk structures,}
	the correction method will always produce more accurate polarization measurements than reduction methods without a model-based correction.
	When the aim of the observation is to perform a qualitative analysis of the data, such as describing the large-scale morphology of disks, a more conventional reduction method may su\ffi ce for a face-on disk such as TW\,Hydrae.
	{\revia However, the loss of polarization signal as displayed between panels a1, a2 \& a3 illustrates that our combination of data from multiple polarimetric cycles will result in very poorly constrained polarimetric intensity measurements.
	More importantly}, when we observe disks at larger inclination, especially in $H$ or $K_s$ band, even qualitative analysis of the data is likely to become skewed when the instrumental polarization e\ff ects are not corrected properly.
	In Paper\,II we illustrate that the shape of $Q_\phi$ and especially $U_\phi$ images of the disk around T\,Cha look much more reliable \citep[i.e. more similar to radiative-transfer model predictions,][]{Pohl:2017}) after applying the correction method.

\section{Recommendations for IRDIS/DPI}
\label{sec:recommend}
	\subsection{Observing strategy to optimize polarimetric performance}
	\label{sec:recstrategy}
	Previous publications \citep[e.g.,][]{Garufi:2017} have demonstrated that the polarimetric mode of IRDIS is a very e\ff ective tool to image the scattering surface of protoplanetary and debris disks. However, the analysis of the TW\,Hydrae data presented in this paper illustrate that the polarimetric e\ffi ciency can be negatively a\ff ected when the instrument con\fix guration is not optimized.
	When using IRDIS/DPI in \fix eld-tracking mode, we recommend the following adjustments to the observing strategy to avoid a loss of polarized signal:

	\begin{itemize}
		\item When no strict wavelength requirements are present and the disk surface brightness is expected to be gray in the NIR:
		 use $J$ band to achieve a ${>}\,90\%$ polarimetric e\ffi ciency, which is nearly independent of the remaining instrumental setup.
		\item However, most young stars are red, causing their disks to scatter more light in $H$ than in $J$ band. 
	If the previous recommendation to use $J$ band cannot be met and the $Y$-, $H$-, or $K_s$-band \fix lters are used, avoid the use of derotator angles 
		$22.5 \lesssim | \theta_\mathrm{der} | \lesssim 67.5^\circ + n \cdot\,90^\circ$, with $n \in \mathbb{Z}$.
	\end{itemize}

	The latter recommendation ensures a polarimetric e\ffi ciency $\geq 70\%$.
	This constraint on $\theta_\mathrm{der}$ can be achieved with two simple steps:

	\begin{enumerate}
	\item Within ESO's Phase 2 Proposal Preparation tool (P2PP or P2),
	split the total observation within an observing block (OB) 
	into parts (templates) where the di\ff erence between 
	$p$ and $a$ does not vary by more than $90^\circ$, resulting in $|\Delta \theta_\mathrm{der}|  < 45^\circ $, because $\theta_\mathrm{der} \propto (a - p)/2$;

		\item For 
		each template,
		the $\theta_\mathrm{der}$ constraint can be determined by \fix nding the average parallactic and altitude
		angles and applying a derotator (position angle) o\ff set of: 
		\begin{equation}
		\mathrm{INS.CPRT.POSANG} =\,{<}\,p-a\,{>} +\,n\,\cdot\,180^\circ,
		\label{eq:req}
		 \end{equation}
		 with $n \in \mathbb{Z}$, and $<>$ indicating the average values.
	\end{enumerate}		

The parallactic and altitude angle can be determined by:
\begin{eqnarray}
	p &=& \arctan \left( \frac{\sin{({HA})}}{\tan{{(\phi)}} \cos{(\delta)} - \sin{(\delta)}\cos({HA}) }  \right ), \\
	a &=& \arcsin \left( \sin{(\delta)}\sin{(\phi)} + \cos{(\delta)} \cos{(\phi)}\cos{(HA)} \right ),  
\end{eqnarray}
where $\phi = -24.6^\circ$ is the latitude of the VLT at the Paranal observatory, $\delta$ is the declination of the star, and
${HA}$ is the Hour Angle of the star = Local Sidereal Time (LST) - Right Ascension (RA).

The required {\revia derotator} o\ff set will be strongly dependent on the exact start of the observation template.
During \textit{visitor mode} observations, where the observer is present at the observatory, the optimal derotator o\ff set can be included in the OBs just before the start of the observations.
During remote \textit{service mode} observations, the observer does not know at what time the observation will start and therefore not what the optimal value will be for INS.CPRT.POSANG.
This problem can be solved by either including a list of LST values and the corresponding derotator position angle o\ff sets to the {README} \fix le of the OB, or by providing LST constraints and a corresponding optimal derotator o\ff set to the OB, preferably around a time when the parallactic angle does not change too much during the observation.

	 When the optimal value for INS.CPRT.POSANG is used, the mean $\theta_\mathrm{der}$ lies at 
	 either $0^\circ$ or $90^\circ$, which orients the derotator horizontal or vertical with respect to the Nasmyth platform.
	This orientation will cause little or no crosstalk and therefore limited loss of 
	polarized signal and a close to ideal polarization angle.

	 \begin{figure}[!h]
   \centering
   \includegraphics[width=0.42\textwidth, trim = 20 10 20 0]{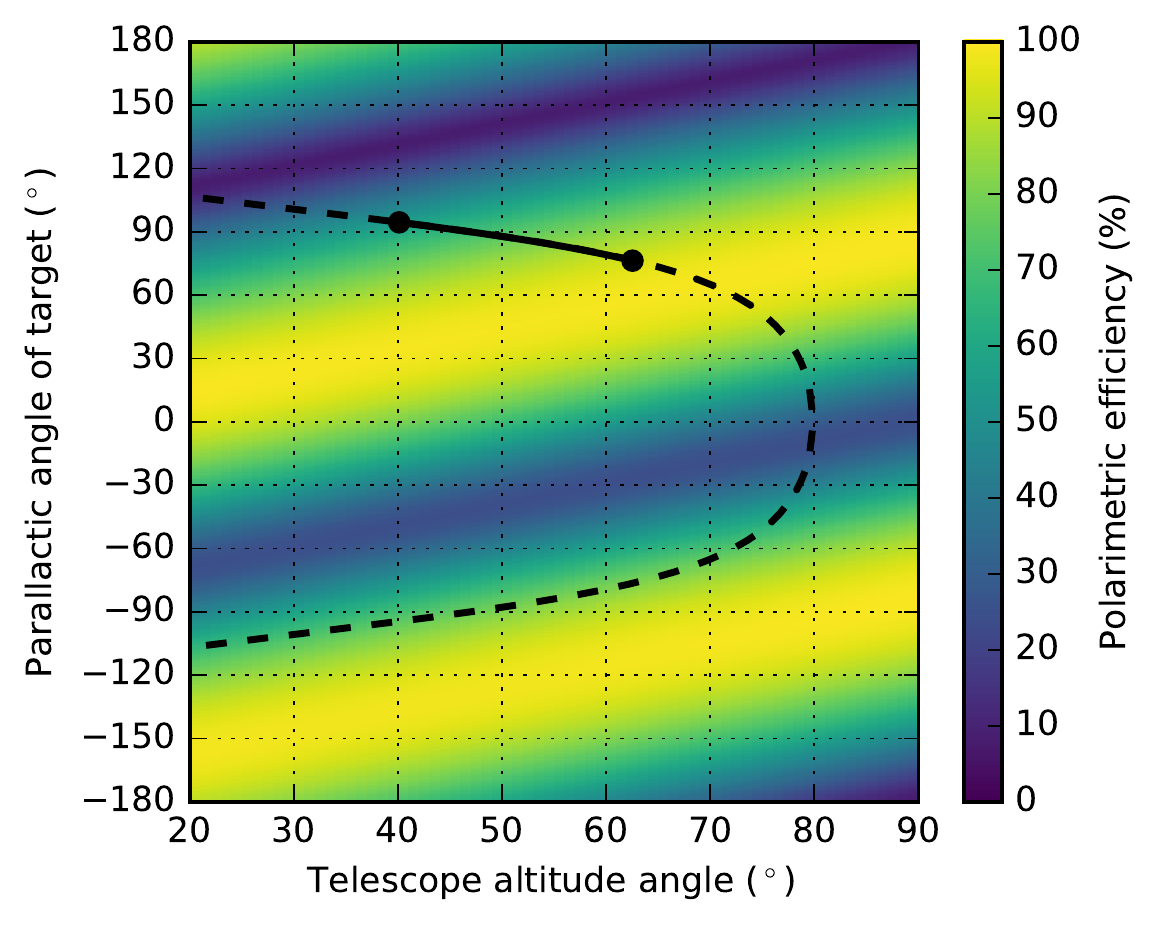}
   \caption{The polarimetric e\ffi ciency in $H$ band as a function of the parallactic angle and altitude angle,
   		when the derotator is in \fix eld-tracking mode without an additional derotator o\ff set. 
		The dashed line shows the parallactic and altitude angle of TW\,Hydrae across the sky, with the solid line showing the angles during the SPHERE observation. 
		We clearly see that using default derotator settings (North up on the detector) will give us mediocre polarimetric e\ffi ciencies.
      \label{fig:effpa}}
    \end{figure}

	Figure~\ref{fig:effpa} shows the polarimetric e\ffi ciency mapped for $p$ and $a$
	without the application of a derotator o\ff set.
	Given the path traveled by TW\,Hydrae during our observations shown with the black solid line 
	(${<}\,p\,{>} \approx 90^\circ, {<}\,a\,{>} \approx 50^\circ$), we could have avoided the loss of polarization signal by 
	providing a derotator position angle o\ff set of $\mathrm{INS.CPRT.POSANG} = 90 - 50 = 40^\circ$,
	rotating the derotator by $20^\circ$.
	We did not apply this derotator o\ff set because we were not aware of the strong crosstalk for IRDIS/DPI at the time of these observations.

	\subsection{SPHERE design upgrades}
	\label{sec:recdesign}

	\subsubsection{HWPs\,1\,\&\,3 to reduce instrumental polarization e\ff ects} 
	\label{sec:hwp123}

	The most thorough way to reduce instrumental polarization e\ff ects for IRDIS would be to introduce two more HWPs in the optical path. A HWP1 at the same location as for ZIMPOL, in between M3 and M4, can keep the {\revia \textit{IP}} induced by both mirrors perpendicular to let them cancel out.
	We then use HWP2 in a slightly di\ff erent fashion: rather than aligning the to-be-measured polarization angles with the analyzers,
	it aligns the desired polarization with the re\fl ection plane of the derotator, as it does for ZIMPOL \citep{Schmid:2018}. 
	Similar to ZIMPOL, this new rotation law for HWP2 requires that we install a third HWP to align the desired polarization angle with the transmision axes of IRDIS' polarizers.
	We recommend to install this HWP3 directly downstream of the derotator, to completely remove all crosstalk.

	We are aware that the diverging beam downstream of the derotator is relatively large, which might make it di\ff icult to produce a HWP that is large enough.  
	The feasibility of this recommendation is yet to be determined. 
	Alternatively, as for ZIMPOL, HWP3 could be installed further down the optical path when the beam starts to become smaller (e.g., between the apodizer wheel and the DTTS beam splitter, before large incidence-angle re\fl ections are encountered by the beam).
	
	\subsubsection{Recoating of the derotator to reduce retardance}

	As we describe in Sect.\,\ref{sec:derot}, due to its unfavorable retardance (nearly $\lambda /4$ in $H$ and $K_s$) the derotator is by far the largest contributor of crosstalk and 
	hence loss of polarimetric e\ffi ciency. 
	Therefore, an alternative recommendation to reduce the loss of polarimetric e\ffi ciency for IRDIS is to recoat the three mirrors of the derotator with a coating that yields a total retardance of the derotator close to $\lambda/2$ in all \fix lters.
	We are currently investigating the feasibility to apply a coating to the derotator that yields ${\sim}\,\lambda/2$ retardance over the very broad wavelength range required by ZIMPOL and IRDIS combined. 
Whether HWP3 is still desired after recoating depends on how close the retardance of the new coating is to the ideal value for the full wavelength range covered by IRDIS.

	\subsubsection{A polarizing beam splitter to increase throughput}
	The throughput of the combination of non-polarizing beam splitter plate + wire-grid polarizers is ${\sim}\,50\%$, as discussed in Sect.\,\ref{sec:analyzers}.
	Replacing the non-polarizing plate with a \textit{polarizing} beam splitter plate will immediately increase the throughput with a factor ${\sim}\,2$.
	An additional result of this upgrade will also be that polarimetry is o\ff ered 'for free' for any observation performed with IRDIS.
	Polarimetry-for-free will allow a substantial boost of the science output of the instrument by serendipitous discoveries of 	polarized circumstellar disks during planet-hunting surveys.
	The observer can choose whether to use full HWP cycles if polarimetry is not the primary objective.
	However, one should always consider at least cycling two HWP angles set $45^\circ$ apart (e.g., $Q^\pm$), especially for DBI to avoid confusing polarized signal for a spectral feature.
	This necessity of the HWP requires that we need to investigate whether inlcuding this optical component a\ff ects the contrast of DBI, classical imaging, but also IFS observations.

	\subsubsection{IRDIS polarimetry \& IFS observations simultaneously}
	An important \fix rst test for the desirability of IRDIS polarimetry-by-default is to do combined IRDIS/DPI + IFS observations. 
	Although this combination is currently not o\ff ered, it does not require any changes in design, only software (new observing templates).
	The instrumental polarization e\ff ects of inserting a dichroic beam splitter in stead of a mirror (to fold the beam towards IRDIS) is expected to be negligible, because it would cancel after the double di\ff erence.
	The largest unknown will be the e\ff ect of the HWP for IFS observations.
	This test is valuable, not only to investigate if a polarizing beam splitter should be installed in IRDIS.
	If the outcome is indeed that IRDIS/DPI observations do not affect the IFS and vice-versa, we again open a new window for increased science output per observation that can be used almost immediately.
	This new mode would be very helpful for substellar companion searches close to stars that are surrounded by disks, or disk-searches while characterizing substellar companions.
	
\section{IRDIS/DPI compared to contemporary AO-assisted imaging polarimeters}
\label{sec:competition}

	{
	In this Section we make a brief comparison between the polarimetric mode of SPHERE/IRDIS and the major contemporary AO-assisted
high-contrast polarimetric imagers operating in the near infrared: GPI and NACO. 
	The designs of both GPI and SPHERE are primarily focused on the minimization of wavefront errors \citep{Macintosh:2014, Beuzit:2019}, 
	which is crucial for the detection of planets at high contrasts and small separation from the central star.
	Polarimetry had a much lower priority in the design choices, 
	which resulted in designs that are suboptimal for the polarimetric performance of both instruments.
	
However, the extreme AO systems of both GPI \citep{Poyneer:2014} and SPHERE \citep{Fusco:2006} turn out to be crucial for their polarimetric imaging modes. 
	Although a detailed comparison between the perfomance of these AO systems and the older generation 
	Nasmyth Adaptive Optics System \citep[NAOS,][]{Rousset:2003hh} of NACO lies outside the scope of this study, 
	the reported performances of these systems allow for some obvious conclusions. 
	The AO systems of SPHERE and GPI can both run at ${\sim} 1$\,kHz and control a high-order DM containing $41\,{\times}\,41$ and $64\,{\times}\,64$ actuators, respectively, while NAOS can control its DM with 185 active actuators at either $444$\,Hz or $178$\,Hz for its visible light or NIR wavefront sensor, respectively.
	The resulting point spread functions reach high Strehl ratios (${\gtrsim}\,90\%$ in $H$-band) and remain very stable for extended periods of time for both SPHERE and GPI, while NACO reaches typical Strehl ratios of 
 ${\gtrsim}\,50\%$ in $K_s$ band and ${\gtrsim}\,30\%$ in $H$ band. 
	Another important characteristic of the three AO systems is the limiting magnitude at which these systems can still operate. 
	These magnitude limits are $I \approx 10$\,mag for GPI \citep{Macintosh:2014}, $R \approx 15$\,mag for SPHERE \citep{Beuzit:2019}, and $V \approx 16$\,mag and $K \approx 14$\,mag for NACO 
\citep{Rousset:2003hh}.
	These limits make SPHERE and NACO particularly well-suited to perform polarimetric imaging observations of relatively faint objects, such as most nearby T\,Tauri stars. 

There are many di\ff erences in the designs of the polarimetric modes of NACO, GPI and SPHERE/IRDIS. 
SPHERE is mounted on the Nasmyth platform of the telescope and employs an internal image derotator to stabilize the \fix eld or pupil. 
NACO is also mounted at the Nasmyth focus, but contrary to SPHERE it is attached to the derotator flange of the telescope support structure \citep{Lenzen:2003iu}. 
This Nasmyth derotator rotates the complete instrument to track either \fix eld or pupil, although pupil tracking is rarely used for the polarimetric imaging mode of NACO. 
GPI is mounted at the telescope Cassegrain focus, which avoids the need of a tertiary telescope mirror. 
It has no image derotator, which allows only pupil stabilized imaging, not \fix eld tracking.
	
	The three instruments contain many internal re\fl ections before the light beam reaches the detector. 
	Because the double di\ff erence removes the $IP$ produced by the optical components downstream of the HWP, the location of the HWP is important for the polarimetric performance of the instruments.
	While HWP2 of SPHERE/IRDIS is situated early in the optical train, directly after the \fix rst internal re\fl ection within the instrument (between M4 and the derotator, see Fig.\,\ref{fig:vltsphere}), 
	GPI and NACO have their HWPs installed after many 
re\fl ective surfaces 
\citep[respectively]{Perrin:2010, Witzel11}.

	As a result of these design choices, the $IP$ of SPHERE originates only from the telescope and M4, and therefore varies with telescope altitude angle (see Table\,\ref{tab:m3m4}). 
The $IP$ of NACO is produced by the telescope and the re\fl ections in the NAOS AO system. 
Because NACO rotates with respect to the telescope in \fix eld-tracking mode, its $IP$ also depends on the telescope altitude angle.
Millar-Blanchaer et al.\,(in prep.) measure the $IP$ of NACO in pupil tracking (most favorable instrument con\fix guration) in $H$ band, 
and determine the $IP$ to be comparable to that of SPHERE/IRDIS at low altitude angles (worst con\fix guration).
The $IP$ of GPI originates primarily from the re\fl ections in the instrument upstream from the HWP. 
\citet{Millar:2016} measure average $IP$ values that are very comparable to SPHERE/IRDIS. 

The crosstalk of SPHERE/IRDIS is predominantly caused by the derotator and HWP2, which can result in a large decrease in polarimetric e\ffi ciency (Sect.\,\ref{sec:derot}). 
Since GPI does not have an image derotator and has all re\fl ections aligned in a single plane, the retardance of the instrument is very small \citep{Millar:2014} compared to SPHERE/IRDIS.
In NACO, the crosstalk is predominantly caused by M3 and the optical components of NAOS. 
Because the HWP is located downstream of these components, the resulting polarimetric e\ffi ciency di\ff ers between measurements of Stokes $Q$ and $U$, when $Q$ is aligned with the optical axes of the polarizing beam splitter. 
\citet{Avenhaus:2014} determine the polarimetric e\ffi ciency of Stokes $U$ to be ${\sim}\,60\%$ relative to Stokes $Q$ for NACO. Although a similar e\ff ect may be expected for GPI due to the location of the HWP, to the best of our knowledge no di\ff erences in polarimetric e\ffi ciency between $Q$ and $U$ have been reported in literature.

	Both \citet{Millar:2016} and \citet{Holstein:2017} show for GPI and SPHERE/IRDIS, respectively, that the polarized contrast at separations ${\gtrsim}\,0.3"$ is dominated by the photon and readout noise and therefore scales with the square root of the exposure time. 
	Both papers report very similar polarized contrasts of $10^{-6} - 10^{-7}$  at a separation of 0.4", 
	while de Juan Ovelar (in prep.) has measured a polarized contrast of ${\sim}\,10^{-6}$ at a separation of 1" for NACO.
	We can therefore conclude that the polarimetric performances of GPI and SPHERE/IRDIS are very similar for relatively bright stars ($R \approx 6$\,mag). 
Due to their extreme AO systems and small di\ff erential wavefront errors, SPHERE/IRDIS and GPI reach much higher polarized contrasts than NACO at sub-arcsecond separations from the star.

\section{Conclusions}
\label{sec:conclude}

	The polarimetric mode of SPHERE/IRDIS {\revia has been very succesful at imaging protoplanetary disks at resolutions and polarized contrasts close to the star 
that were not attainable with the previous generation of polarimetric imagers.} 
	Because the design was mainly driven by non-polarimetric requirements, its performance is strongly dependent on the observing strategy,
	as we have illustrated with the observations of TW\,Hydrae.
	When the observing strategy is not optimized, polarimetric crosstalk can cause the e\ffi ciency 
	to drop towards ${\sim}\,5\%$ in $H$ and $K_s$ band; polarimetric e\ffi ciency remains above $54\%$ in 
	$Y$ band
	and above $89\%$ in $J$ band.
	Low polarimetric e\ffi ciency means that we lose polarization signal, which is what we aim to detect in DPI mode.
	Crosstalk also causes a {\revia polarization angle} o\ff set 
up to ${\sim}\,30^\circ$ in $H$- and more in $K_s$.
	We have demonstrated that the polarimetric instrument model described in Paper\,II can be used to explain and correct for the variations in
	polarimetric e\ffi ciency and polarization angle o\ff set due to crosstalk observed in the TW\,Hydrae data.

	The work presented in Papers\,I\,\&\,II show that instrumental polarization e\ff ects are significant but also 
	that this a-posteriori work allows a very high quality data product, which is above expectations given the loose constraints on the design requirements.
	Optimal results can be obtained from IRDIS/DPI observations when two important considerations are taken into account:
	1) Adjust the observating strategy beforehand as described in Sect.\,\ref{sec:recstrategy} to minimize a decrease in 
	polarimetric e\ffi ciency. 
	2) Apply the correction method described in Paper\,II, and included in the IRDAP pipeline, to correct the data for instrumental polarization e\ff ects.


\begin{acknowledgements}
A significant part of this work was performed when JdB, RGvH and
JHG were a\ffi liated to ESO. JdB and RGvH thank ESO for the studentship 
at ESO Santiago during which this project was started. Many thanks
go out to the SPHERE team and the instrument scientists and operators of the
ESO Paranal observatory for their support during the calibration measurements.
The research of JdB and FS leading to these results has received funding from
the European Research Council under ERC Starting Grant agreement 678194
(FALCONER).
SPHERE is an instrument designed
and built by a consortium consisting of IPAG (Grenoble, France), MPIA (Heidelberg,
Germany), LAM (Marseille, France), LESIA (Paris, France), Laboratoire
Lagrange (Nice, France), INAF - Osservatorio di Padova (Italy), Observatoire
de Genève (Switzerland), ETH Zurich (Switzerland), NOVA (Netherlands), ONERA
(France), and ASTRON (Netherlands) in collaboration with ESO. SPHERE
was funded by ESO, with additional contributions from the CNRS (France),
MPIA (Germany), INAF (Italy), FINES (Switzerland) and NOVA (Netherlands).
SPHERE also received funding from the European Commission Sixth and Seventh
Framework Programs as part of the Optical Infrared Coordination Network
for Astronomy (OPTICON) under grant number RII3-Ct-2004-001566 for
FP6 (2004-2008), grant number 226604 for FP7 (2009-2012), and grant number
312430 for FP7 (2013-2016).
\end{acknowledgements}

\bibliography{ref160504}   
\bibliographystyle{aa}   
\appendix
\section{Tracking laws for HWP2 and the derotator}
\label{sec:tracklaw}

\subsection{Field tracking}
\subsubsection{The derotator}

Field tracking is the default setting for the polarimetric imaging mode of IRDIS.
In this setting, the derotator control law keeps the image with north up on the detector, 
(except for the true-north o\ff set described in Sect.\,\ref{sec:twhpdi})
which is given by:
\begin{equation}
\theta_{\mathrm{der}} = \frac{1}{2}\left(-p + a\right) + \frac{1}{2}\eta,
\end{equation}
with $\theta_{\mathrm{der}}$ the derotator angle, $p$ the astronomical object's parallactic angle (FITS header keyword: TEL.PARANG.START), and $a$ the altitude angle of the Unit Telescope (TEL.ALT). The user-de\fix ned position angle o\ff set $\eta$ of the image (INS4.DROT2.POSANG) can be altered by changing the value of INS.CPRT.POSANG in the Observing Block (OB), as described in Sect.\,\ref{sec:recommend}. The header value of the derotator angle is computed as: 
\begin{equation}
\mathrm{INS4.DROT2.BEGIN} = \theta_{\mathrm{der}} + n\cdot360^\circ,
\label{eq:derotator_header}
\end{equation}
with $n \in \mathbb{Z}$. 

\subsubsection{The half-wave plate}

\noindent The HWP2 control law for \fix eld-tracking with IRDIS/DPI (implemented in March, 2015) keeps the polarization direction aligned with the analyzers, and is given by:   
\begin{eqnarray}
\theta_{\mathrm{HWP}} &=& -p + a + \frac{1}{2}\left(\eta + \gamma\right) +\theta_\mathrm{HWP}^s \\
 				   &=& 2\theta_{\mathrm{der}} + \frac{1}{2}\left(\gamma -\eta\right) + \theta_\mathrm{HWP}^s,
\end{eqnarray}
with $\theta_{\mathrm{HWP}}$ the HWP angle, and $\gamma$ a position angle o\ff set of the linear polarization direction due to a user-de\fix ned HWP2 o\ff set (INS4.DROT3.GAMMA, to be changed by adjusting SEQ.POL.OFFSET.GAMMA in the OB). $\theta_\mathrm{HWP}^s$ is the switch angle used during HWP cycles as described in Sect.\,\ref{sec:UTCPI}. 
When SEQ.IRDIS.POL.STOKES is set to "QU" in the OB, $\theta_\mathrm{HWP}^s$ cycles through the angles $0^\circ, 45^\circ, 22.5^\circ$ and $67.5^\circ$, but only through $0^\circ$ an $45^\circ$ if SEQ.IRDIS.POL.STOKES is set to "Q". 
The latter setting is not recommended for polarimetric measurements. 
The header value of HWP2 angle is computed as: 
\begin{equation}
\mathrm{INS4.DROT3.BEGIN} = \theta_{\mathrm{HWP}} + 152.15^\circ + n\cdot180^\circ,
\label{eq:drot3}
\end{equation}
with $n \in \mathbb{Z}$. 

\subsection{Pupil tracking}
\subsubsection{The derotator}
\label{sec:ptderot}
The derotator control law for pupil-tracking with IRDIS keeps the pupil \fix xed, which causes the image to rotate over the detector with the parallactic angle. This tracking law is given by:
%
\begin{equation}
\theta_{\mathrm{der}} = \frac{1}{2}(a + \mathrm{PUPIL_{offset}}),
\label{eq:theta_der_pupil}
\end{equation}
with $\mathrm{PUPIL_{offset}} = 135.99\pm0.11^\circ$ (SPHERE Manual) the position angle o\ff set of the image required to align the telescope pupil with the spider mask in the Lyot stop within IRDIS, in order to mask the di\ff raction pattern caused by the M2 support structure (spiders). 
The header value of the derotator angle is computed as: 
\begin{equation}
\mathrm{INS4.DROT2.BEGIN} = \theta_{\mathrm{der}} - \frac{1}{2}\mathrm{PUPIL_{offset}} + n\cdot360^\circ,
\end{equation}
with $n \in \mathbb{Z}$. 
This means that the header value of the derotator angle does \textit{not} include $\mathrm{PUPIL_{offset}}$, and therefore does not represent the true derotator angle as is the case with the header value in \fix eld-tracking (Equation~\ref{eq:derotator_header}). \\

\subsubsection{The half-wave plate}

The {pupil-tracking} law for HWP2 (implemented in January 2019) keeps the polarization direction aligned with the analyzers. 
Therefore, we can simply add up the $Q$- and $U$-images after software-derotating them (the conventional data-reduction method). 
This way the polarization direction is also kept constant during integration and does not smear 
(however, the image, especially at large separations from the star, 
does smear out on the detector during long integrations due to rotation of the \fix eld with parallactic angle).
The following HWP2 control law is implemented:
%
\begin{equation}
\theta_{\mathrm{HWP}} = -\frac{1}{2}p + a + \frac{1}{2}\left(\eta + \gamma + \mathrm{PUPIL_{offset}}\right) +  \theta_\mathrm{HWP}^s.
\label{eq:theta_hwp_pupil}
\end{equation}
The header value of HWP2 can still be determined using Eq.\,\ref{eq:drot3}.

\section{Data pre-processing}
\label{sec:reduce}

\begin{figure*}[!h]
   \centering
   \includegraphics[width=\textwidth, trim = 20 0 0 20]{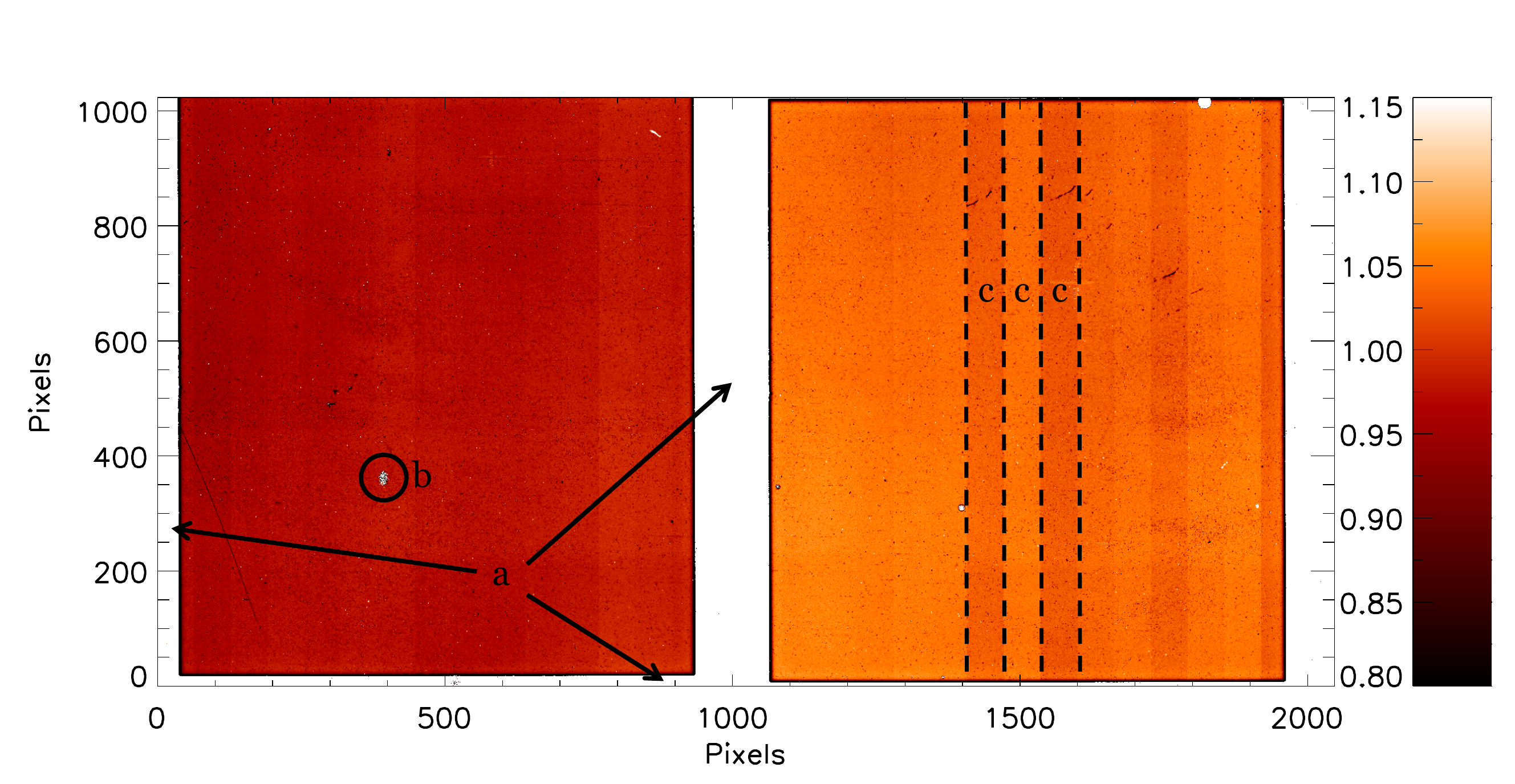}
   \caption{
	Normalized IRDIS master-\fl at image in $H$ band. The illuminated detector halves are surrounded by the 
	\fix eld-mask \{\textbf{a}\}. 
	The masked regions (and all pixels with value $< 0.1$) are set to $1000$ and shown as white in this image.
	This enhances several clusters of dead and hot pixels (e.g., \{\textbf{b}\} around [x,y] $\approx$ [400,400]).
	A time varying read-out signature of the detector is visible as columns of 64 pixels wide (e.g., \{\textbf{c}\}). 
     \label{fig:flat}}
    \end{figure*}

	Raw IRDIS frames consist of $2048$ by $1024$ pixels in the $x$ \& $y$ directions, respectively.
	The two beams separated by the beam splitter (see Sect.\,\ref{sec:desirdis}) are centered roughly on the left and right detector halves,
	while a \fix eld mask avoids leakage of signal from one half to the other.

	We median combine all dark + background observations taken with DIT = 16\,s to create a master-dark image.	
	From internal light source (\fl at-\fix eld) measurements, we create a (dark-subtracted) master-\fl at image: 
	we take median over two regions of $800\,{\times}\,800$ centered on the left half 
	([$x,y$] = [512, 512]) and the right half ([$x,y$] = [1536,512])
	and use this value to normalize.
	To avoid emphasizing dead pixels or pixels masked by the \fix eld mask, we change all pixel values $< 0.1$ into $1000$.
	The \fix nal master-\fl at is shown in Fig.\,\ref{fig:flat}.
	Note that the master-\fl at still contains a vertical read-out pattern (features c in Fig.\,\ref{fig:flat}). 
	It is di\ffi cult to distinguish between the \fl at signal and this pattern, and polarimetry requires that we maintain the di\ff erent transmission ratios of the two detector halves in our master-\fl at (hence the single normalization earlier).
Therefore, we have chosen to leave the master-\fl at uncorrected for this pattern and remove these read-out columns at a later stage in the post-processing (Sect.\ref{sec:twhpdi}).
	 
	Preferably, to correct for background signal we would have used 'SKY' observations: 
	observations taken with the same DIT as the science (or 'OBJECT') frames, 
	recorded on-sky with the star moved out of the detector FOV. 
	Such SKY images contain background emission from possible interstellar origin (e.g., from nearby nebulosity), 
	the earth atmosphere, the optical system and the detector dark current.
	Because there were no SKY images recorded for these observations, we corrected for instrument background 
	emission by subtracting the master-dark from each science frame recorded of TW\,Hydrae. 
	Subsequently, we divide the science frames by the master-\fl at to correct for \fl at-\fix eld errors.
	Next, we crop the frames to separate the left ($1 \leq x \leq 1024$) from the right ($1025 \leq x \leq 2048$) detector half.
	
	To align the central star with the center of the image one could use the 'CENTER' frames, where SPHERE's 
	DM has created satelite spots around the star-center.
	However, because alignment between individual frames is a crucial element of PDI 
	we opted for an alternative multi-stage approach.
	We find the centers of the star in both left and right frame by cross-correlation with a two-dimensional Mo\ff at function, 
	where the central pixels within a radius of six pixels are set to zero
	to account for the coronagraph mask.
	We determine $I$ according to Eq.~\ref{eq:imeas}, and median combine the images of all HWP cycles 
	to create a template image. 
	To enhance our frame-to-frame alignment, we repeat the cross-correlation for all frames, this time using the template image.
	Due to irregular diffraction paterns close to the edge of the coronagraph mask, 
	we cannot guarantee with the two centering steps described above that the star-center lies at the center of the image. 
	However, it does place the star close to the center of the image, and accurately alligns all frames with respect to each other.
	The latter is crucial to perform the single and double-di\ff erence subtractions.
	A more advanced centering is performed at a later stage.

\end{document}